\documentclass[usegraphicx,useAMS]{mn2e}
\usepackage{subfigure}

\title[Abundances in Blanco~1]
  {Elemental abundances in the Blanco~1 open cluster}
\author[A. Ford et al.]
  {A. Ford$^{1}$, R.D.~Jeffries$^{2}$ and B. Smalley$^{2}$\\
$^{1}$CSPA/SPME, Building 28M, Monash University, VIC 3800, Australia\\ 
$^{2}$Astrophysics Group, School of Chemistry and Physics, Keele
  University, Keele,
      Staffordshire ST5 5BG, United Kingdom}
\date{Submitted August 2005}

\pagerange{\pageref{firstpage}--\pageref{lastpage}} \pubyear{2004}

\def\LaTeX{L\kern-.36em\raise.3ex\hbox{a}\kern-.15em
    T\kern-.1667em\lower.7ex\hbox{E}\kern-.125emX}

\begin{document}

\label{firstpage}

\maketitle

\begin{abstract}
High resolution spectroscopy is used to determine the detailed chemical
abundances of a group of eight F- and G-type stars in the young open
cluster Blanco~1. An average [Fe/H] of $+0.04$ $\pm0.02$ (internal error)
$\pm 0.04$ (external error) is found,
considerably lower than a previous spectroscopic estimate for this
cluster. The difference is due mainly to our adoption of significantly
cooler temperatures which are consistent with both photometric and
spectroscopic constraints. Blanco~1 exhibits sub-solar [Ni/Fe]
($-0.18$ $\pm0.01$ $\pm0.01$), [Si/Fe] ($-0.09$ $\pm0.02$ $\pm0.03$),
[Mg/Fe] ($-0.14$ $\pm 0.02$ $\pm0.03$)
and [Ca/Fe] ($-0.09$ $\pm0.03$ $\pm0.03$); ratios which are not observed among
nearby field stars.  The material from which Blanco~1 formed may not
have been well mixed with interstellar matter in the galactic disc,
which tallies with its current location about 240\,pc below the
galactic plane. A simultaneous deficit of Ni and alpha elements with
respect to Fe is hard to reconcile with most published models of yields
from supernovae of types~Ia and II. The revised abundances for Blanco~1
indicate that overall radiative opacities in its stars, and hence
convective zone properties at a given mass, are similar to those in the
Pleiades at approximately the same age. This can explain a previous
observation that the Li depletion patterns of G- and K-type stars in
the two clusters are indistinguishable. The lower overall metallicity
of Blanco~1 now make it less attractive as a target for discovering
transiting, short period exoplanets.

\end{abstract}


\section{Introduction}
Open clusters are excellent laboratories for testing our understanding
of stellar structure. Their numerous stars share common ages and
distances, reducing many uncertainties associated with field-star
studies. Abundances for elements other than Fe and Li are rarely
available in open clusters, yet these have a profound bearing on
stellar structure calculations.

A case in point is the Blanco~1 cluster which has an age similar to, or
a little younger than, the Pleiades (50-100\,Myr -- Perry, Walter \&
Crawford; Panagi et al. 1994).  Edvardsson et al. (1995, hereafter E95)
claimed [Fe/H]\,=$+0.23$ for the cluster on the basis of spectroscopy
of several F stars. E95 discussed this high metallicity in terms of the
unusual location of Blanco~1. The cluster is 240\,pc below the galactic
plane and may have crossed the plane on one or more occasions.  The
apparent metal-rich status of the cluster has lead to a number of
investigations that have sought to isolate the composition dependence
of various physical phenomena. Pillitteri et al. (2003, 2004) have used
Blanco 1 to determine whether metallicity influences the coronal X-ray
losses from low-mass stars with convective envelopes. Blanco~1 may well
be a fruitful location to search for transiting exoplanets, given the
established relationship between stellar metallicity and the frequency
of short-period exoplanets around field stars (e.g. Santos, Israelian
\& Mayor 2004 and references therein).

Jeffries \& James (1999, hereafter JJ99) found that the Li depletion
pattern with $T_{\rm eff}$ among the G/K stars of Blanco~1 could not be
distinguished from the similarly aged Pleiades, which has
[Fe/H]$=-0.03$ (Boesgaard \& Friel 1990). This result contradicts the
strong metallicity dependence predicted by models of pre-main-sequence
(PMS) Li depletion, implying that some unknown mechanism inhibits PMS
Li depletion in the Blanco~1 stars and that some non-convective mixing
process operates in main sequence stars to ensure that Li-depletion in
a metal-rich ZAMS cluster like Blanco~1 could approach 
that seen in the Hyades after 700\,Myr.

A possible escape route for the ``standard'' PMS models is if elements,
other than Fe, that also form a significant source of opacity in PMS
stars (particularly oxygen), are {\it underabundant} in the Blanco~1 stars
compared to a solar mixture. This would mean that, overall, the
opacities in the outer envelope of the Blanco~1 stars could be similar to
those in the Pleiades, leading to similar levels of PMS Li depletion
(see Piau \& Turck-Chi\`eze 2002).

A further issue to address is whether the metallicity determination of
E95 can be confirmed. It seems
most likely that several of the stars analysed by E95 were {\it{not}}
cluster members and that as chemical composition was used as a
memership discriminator, the overall result might have been biased.
Additionally, the temperature scale used in E95 for their atmospheric
analyses was considerably hotter than other commonly used calibrations,
resulting in a higher overall metallicity estimate (JJ99). Westerlund
et al. (1988) found that the cluster was slightly metal deficient on
the basis of their Stromgren photometry.

In this paper we analyse high resolution echelle spectroscopy of a
sample of candidate F and G stars in Blanco~1. The aims of our study
are to confirm whether Blanco~1 really is a very metal-rich open
cluster and to find individual abundances for a large range of elements
including those important in determining stellar structure such as
oxygen, silicon and magnesium.

\section{AAT Spectroscopy}

\subsection{Target selection}

Targets were selected to be F- or G-type cluster candidates (to
facilitate a differential abundance analysis with respect to the Sun)
with low projected rotational velocity and a radial velocity consistent
with cluster membership. Four stars were chosen on the basis of
information published in JJ99 and a further four from information
provided by E95 and J.-C. Mermilliod. (priv. comm).  Photoelectric (or
photgraphic) Johnson $V$-band photometry is available for all these
targets from Westerlund et al. (1988) or de Epstein \& de Epstein
(1985), and $JHK$ photometry is available from the 2MASS catalogue
(Cutri et al. 2003).

\subsection{Observations}

High-resolution spectra were collected on 2002 August 25-27 using the
3.9-m Anglo-Australian Telescope (AAT) and a MITLL3 CCD. The University
College London Echelle Spectrograph (UCLES) was utilised with the E31
grating to obtain data with R$\simeq$50000 around a central wavelength of
6463{\AA} in order 88.  The usual calibration frames were collected,
including a reflected solar spectrum (using the Moon) and radial
velocity standards. Several exposures were taken of each object,
resulting in total integration times of 2400 to 8000\,s and
signal-to-noise ratios (SNR) of 70-100 per 0.07{\AA} pixel in the
central orders.

\subsection{Data Reduction}

The data were reduced using the Starlink {\sc{echomop}} (Mills et
al. 1997) package, which
included bias subtraction, flat fielding, subtraction of scattered
light and wavelength calibration.  Radial velocities (RVs) were
measured by cross-correlation of data with wavelengths 5570--5680{\AA},
6050--6160{\AA}\ and 6350--6450{\AA}, which contained a number of
sharp, neutral metal lines and little telluric contamination. The
IAU RV standard star used was HR 6349, with an assumed heliocentric velocity
of -17.4\,km\,s$^{-1}$.  The calculated RVs are presented (together with
other stellar parameters) in Table~\ref{table-params}.  RV
uncertainties quoted are just the standard deviation from 
cross-correlation of the three separate wavelength ranges. There are
additional systematic external errors that probably amount to $\simeq
1$\,km\,s$^{-1}$.

\begin{table*}
\caption{Derived atmospheric parameters for sample stars. Identifiers
in {\bf{bold}} are those used throughout this paper. W identifiers are
taken from Westerlund et al 1988; ZS identifiers are those from de
Epstein \& de Epstein (1985). Effective temperatures derived from
photometry and from spectroscopy (see text) are listed.}
\begin{tabular}{llllllrlrr}
\hline
Object & ZS & $V-K$ & T$_{V-K}$ & T$_{\rm{spec}}$ & $\log g$ & [Fe/H] & $\xi$ & RV & $v \sin i$ \\
       &    &       & K & K    &          &        & km\,s$^{-1}$ & km\,s$^{-1}$ & km\,s$^{-1}$  \\
\hline
{\bf{W64}}   & ZS102&1.77& 5430 & 5440 & 4.60 & $-0.04\pm0.02$  & 1.63 & 5.1$\pm$0.4 &  5.8$\pm$1.4  \\
--           & {\bf{ZS58}} &1.74& 5480 & 5650 & 4.56 & $-0.05\pm0.02$ & 2.21 & 3.0$\pm$1.4 & 12.0$\pm$1.2  \\
--           & {\bf{ZS141}}&1.57& 5730 & 5750 & 4.53 & $+0.09\pm0.02$  & 1.25 & 4.8$\pm$1.4 &  4.8$\pm$1.8  \\
{\bf{W113}}  & ZS182&1.41& 5980 & 5900 & 4.50 & $+0.09\pm0.02$  & 1.15 & 4.8$\pm$1.2 &  6.5$\pm$1.3 \\
{\bf{W60}}   & ZS101&1.06& 6600 & 6450 & 4.38 & $+0.11\pm0.03$  & 1.66 & 5.0$\pm$1.1 & 24.4$\pm$2.9  \\
{\bf{W63}}   & ZS99 &1.10& 6515 & 6480 & 4.38 & $+0.05\pm0.02$  & 1.28 & 2.5$\pm$0.7 &  3.1$\pm$1.9  \\
{\bf{W8}}    & --   &1.13& 6470 & 6540 & 4.38 & $+0.09\pm0.02$  & 1.68 & 3.4$\pm$0.3 & 13.4$\pm$2.2  \\
{\bf{W38}}   & ZS48 &1.14& 6450 & 6550 & 4.37 & $-0.03\pm0.02$  & 1.48 & 4.5$\pm$0.5 &  9.4$\pm$2.3  \\
Sun          & --   & &       & 5780 & 4.44 &  0.00 & 1.21 & 0.0 & 2 \\
\hline

\end{tabular}
\label{table-params}
\end{table*}

\section{Analysis}

Data analysis was performed using {\sc{uclsyn}} (Smith 1992; Smalley,
Smith \& Dworetsky 2001), a spectrum synthesis package. Kurucz 1-D,
homogeneous, LTE, {\sc{atlas9}} (Kurucz 1993) model atmospheres with
the mixing-length theory of convection ($\alpha$=1.25) without
overshooting (Castelli, Gratton \& Kurucz 1997) were used.

\subsection{Atmospheric Parameters}
\label{params}

Atmospheric parameters were determined using the spectroscopy. An
initial estimate of $T_{\rm eff}$ was obtained from the photometry and
this leads to a $\log g$ estimate from a 100\,Myr isochrone taken from
the evolutionary models of Schaller et al. (1992). We then used
measurements of the equivalent widths (EWs) of Fe\,{\sc i} lines (see
section~\ref{iron}) to estimate a microtubulence using the Magain
(1986) method. An ``ionization balance'' locus in the $T_{\rm eff}-\log
g$ plane was then found by demanding agreement in the Fe abundances
determined using lines arising from both neutral and ionized species.
The intercept of this locus with the model isochrone gave a new $T_{\rm
  eff}-\log g$ estimate and the process was iterated using an atmosphere
with a metallicity [M/H] equal to [Fe/H]. 

This process quickly converged, yielding the parameters quoted in
Table~\ref{table-params}.  Quoted uncertainties in [Fe/H] are solely
internal errors and are discussed further in section~\ref{errors}.
Projected rotation velocities ($v \sin i$) were simultaneously
estimated during this process by synthesising the Fe lines with
differing $v \sin i$ and minimising chi-squared.  The errors quoted for
$v \sin i$ in Table~\ref{table-params} are the standard deviation about
the average rotational velocity value for each object, based on between
10 and 32 lines.

The atmospheric parameters in Table~\ref{table-params} were used as
the basis for determining the abundances of several chemical elements.

\subsection{Iron}
\label{iron}

Lines for study were taken from the list included in Reddy et al. (2003
- hereafter R03), and also from E95.
Those lines with $\log gf$ measurements rated B or above in the NIST
Atomic Spectra Database were measured in the solar spectrum, and
abundances determined. The broadening parameters of Barklem, Piskunov \&
O'Mara (2000)  were used for the van der Waals broadening where
available. Adopting $T_{\rm eff}=5780$\,K, $\log g = 4.44$ and a
microturbulence of 1.21\,km\,s$^{-1}$, an average solar abundance of $A$(Fe)$=
7.44\pm0.01$ was determined. This value was then assumed in order to obtain
astrophysical $\log gf$ values for lines with laboratory $\log gf$
values of lower accuracy.  

The EWs were determined by integration of a fitted synthetic spectrum
for a total of 34 Fe\,{\sc{i}} and 11 Fe\,{\sc{ii}} lines in our
reflected solar spectrum. The lines used, their assumed atomic
parameters and measured solar EWs are included in
Table~\ref{table-lines}. We then measured the EWs of as many of these
lines as possible in the Blanco~1 targets, using synthetic spectra and
atmospheric parameters as described in section~\ref{params}. These
measurements are also listed (for Fe and all the other elements
discussed in this paper) in Table~\ref{table-lines}, along with all the
lines of other elements measured in this paper (see below). A number of
lines could not be successfully measured in some of the Blanco~1
targets either as a result of intrinsic weakness, blending with other
lines (which is a function of $v \sin i$) or cosmic ray
contamination. Mean abundances were calculated for each star, based on
estimates of the differential [Fe/H] on a line-by-line basis.

\begin{table*}
\caption{Atomic data and equivalent width measurements}
\label{table-lines}
\begin{tabular}{lllrrrrrrrrr}
\hline
          
Wavelength &$\chi_{p}$& log gf& \multicolumn{9}{c}{Equivalent widths (m\AA)} \\
 \AA       & eV   &       & Sun & W113& W38 & W60 & W63 & W64 &  W8 &ZS141&ZS58 \\
\hline
\multicolumn{3}{c}{Li\,{\sc i}} \\						
\hline
6707.760 & 0.000 & -0.002 & --  &153.1& 61.4& 92.3& 83.4&156.2& 25.2&136.0&228.8\\  
6707.910 & 0.000 & -0.299 & \multicolumn{9}{c}{blended with the above} \\
\hline
\multicolumn{3}{c}{C\,{\sc i}}\\
\hline
5380.320 & 7.685 & -1.679 & 19.9& 16.6& 38.7& 34.9& 29.8& 13.0& 47.3& 20.0& 15.2\\
6587.620 & 8.537 & -1.119 & 13.0& 10.2& 28.5& 31.5& 20.3&  5.6& 32.5&  4.2&     \\
7116.960 & 8.647 & -0.886 & 16.8&     & 34.6& 28.7& 26.7&  5.6& 45.7& 14.8&     \\ 
\hline
\multicolumn{3}{c}{O\,{\sc i}}\\
\hline
7771.944 & 9.146 &  0.369 & 76.2& 85.7&150.0&174.1&139.4& 53.6&174.0& 69.9&     \\
7774.166 & 9.146 &  0.223 & 66.0& 68.8&139.3&118.3&119.9& 41.2&139.3& 59.8& 62.8\\
7775.388 & 9.146 &  0.002 & 50.1& 56.5&112.9&116.7& 99.6& 38.8&119.6& 54.4& 48.8\\
\hline
\multicolumn{3}{c}{Mg\,{\sc i}}\\
\hline
5711.088 & 4.346 & -1.630 &114.6&105.0& 69.3& 81.8& 71.9&130.8& 81.1&118.9&123.7\\
6318.717 & 5.108 & -1.813 & 63.7& 53.0& 28.1&     & 27.4& 75.4&     &     & 87.8\\
7657.603 & 5.108 & -1.224 &104.5& 91.0& 59.1& 61.9& 59.4&107.5& 63.9&104.2&142.9\\
7691.553 & 5.753 & -0.727 &136.8&125.5& 79.3& 93.7& 83.3&141.8& 93.9&170.6&118.7\\
\hline
\multicolumn{3}{c}{Si\,{\sc i}}\\
\hline
6145.016 & 5.616 & -1.453 & 41.2&     & 25.2& 36.5& 29.9& 32.3& 36.8& 44.4& 39.4\\
6721.848 & 5.863 & -1.172 & 50.6& 51.6& 35.3& 38.6& 38.6& 43.7& 40.0& 51.9& 48.7\\
6741.628 & 5.984 & -1.595 & 18.9&     & 10.5&     & 15.7& 12.0& 17.7& 16.2& 15.1\\
6848.580 & 5.863 & -1.703 & 18.7& 16.2& 15.2&     &     & 14.7& 13.6& 13.5&     \\
7405.772 & 5.614 & -0.683 & 99.4& 92.0& 74.6&     & 71.3& 87.5& 87.6& 92.6& 82.6\\
7799.996 & 6.181 & -0.777 & 70.0& 52.2& 31.8&     & 36.5& 52.1& 45.2& 55.3& 42.3\\
7918.384 & 5.954 & -0.642 & 96.9& 91.7& 65.3&     & 65.6& 85.1& 61.5& 87.7& 87.8\\
\hline
\multicolumn{3}{c}{S\,{\sc i}}\\
\hline
6052.674 & 7.866 & -0.568 & 13.5& 13.1& 21.5& 24.2& 23.3&     & 27.9&     & 23.6\\ 
6757.007 & 7.866 & -1.481 & 18.4& 14.2& 31.8& 35.7& 32.4& 11.8& 44.2& 17.1& 18.1\\
6757.171 & 7.866 & -0.417 &   \multicolumn{9}{c}{blended with the above}\\
\hline
\multicolumn{3}{c}{Ca\,{\sc i}}\\
\hline
6166.439 & 2.521 & -1.116 & 76.3& 68.8& 39.3& 38.5& 43.3& 89.0& 48.1& 72.6& 87.5\\ 
6169.042 & 2.523 & -0.771 &101.4&103.4& 64.4& 65.7& 70.3&126.5& 77.0&108.9&130.1\\
6455.598 & 2.523 & -1.304 & 60.1& 59.0& 28.4&     & 34.6& 71.3& 34.9& 57.3& 79.9\\
\hline
\multicolumn{3}{c}{Ti\,{\sc i}}\\
\hline
5219.710 & 0.021 & -2.141 & 28.3& 23.2&     &     &     & 49.2&     & 32.2& 43.3\\
5866.452 & 1.067 & -0.743 & 48.9& 40.5& 20.1& 13.8& 13.5& 67.1& 16.0& 44.4& 68.5\\
6091.174 & 2.267 & -0.299 & 16.0&  8.7&     &     &     & 22.8&     &     & 23.9\\
6126.217 & 1.067 & -1.273 & 23.2& 23.1&     &     &  5.2&     &  8.1& 28.7& 39.3\\
6258.104 & 1.443 & -0.325 & 53.4& 44.3& 16.8&     & 19.5& 66.1& 23.9& 56.0& 68.0\\
\hline
\multicolumn{3}{c}{Fe\,{\sc i}}\\
\hline
5778.455 & 2.590 & -3.397 & 20.9& 20.8&     &     &     & 32.1&     & 24.8&     \\
5784.657 & 3.400 & -2.465 & 29.4&     &     &     & 12.4&     &     &	  &     \\
5809.217 & 3.884 & -1.589 & 48.3& 46.6& 26.1&     & 26.7& 63.2& 26.9& 54.4& 60.8\\
5852.217 & 4.549 & -1.100 & 42.0& 39.4&     &     & 24.2& 49.9& 24.6& 43.5& 48.3\\
5855.090 & 4.608 & -1.436 & 21.7&     &     &     & 11.7& 24.1&     & 21.0&     \\
5858.779 & 4.220 & -2.159 & 13.2& 10.2&     &     & 	& 17.6&     &	  &     \\
5859.600 & 4.549 & -0.610 & 74.4& 79.4& 48.0& 68.5& 51.9& 90.9& 59.5& 78.3& 82.2\\
5862.353 & 4.549 & -0.250 & 89.8& 93.7& 66.4& 82.2& 69.8&109.7& 75.4&	  &111.1\\
5956.692 & 0.859 & -4.506 & 54.9& 55.2& 11.2&     &	  & 74.9&     &	  & 72.0\\
6027.050 & 4.076 & -1.170 & 64.8& 66.5& 38.6&53.9 & 42.4& 71.4& 46.5& 63.6& 73.4\\
\hline
\end{tabular}
\end{table*}

\begin{table*}
\begin{tabular}{lllrrrrrrrrr}
\hline
          
Wavelength &$\chi_{p}$& log gf& \multicolumn{9}{c}{Equivalent widths (m\AA)} \\
 \AA       & eV   &       & Sun & W113& W38 & W60 & W63 & W64 &  W8 &ZS141&ZS58 \\
\hline
\multicolumn{3}{c}{Fe\,{\sc i}}\\
\hline
6151.617 & 2.176 & -3.248 & 48.2& 41.9& 13.3&     & 19.6&     &     & 56.7& 59.8\\
6159.368 & 4.608 & -1.779 & 12.6&     &	    &	  &  7.1&     &  7.5& 14.0&     \\
6165.361 & 4.143 & -1.460 & 44.7& 42.9&	    &	  & 23.8& 49.5& 28.7& 45.8&     \\
6173.340 & 2.223 & -2.790 & 67.0& 72.5& 33.1& 43.7& 41.4& 90.0& 46.6& 80.8& 90.5\\
6200.314 & 2.609 & -2.285 & 71.0& 78.1& 46.3&	  & 45.2& 98.7& 59.7& 87.7&111.0\\
6213.429 & 2.223 & -2.505 & 84.5& 81.1& 47.2&	  & 53.2&120.7& 53.2&	  &     \\
6240.645 & 3.233 & -3.207 & 50.1&     &	    &	  & 19.3& 66.3&	    &	  &     \\
6265.131 & 2.176 & -2.522 & 82.5& 83.3& 52.3&	  &	&115.8& 56.7& 94.5&122.3\\
6358.693 & 0.859 & -3.963 & 80.4& 80.5& 32.0& 41.0&	&107.2& 38.5& 95.9&     \\
6436.411 & 4.186 & -2.360 &  9.7&     &	    &	  &	& 12.8&	    & 12.1&     \\
6481.869 & 2.279 & -2.852 & 61.2& 62.0&	    & 42.7&	& 81.9& 43.4& 68.6& 86.8\\
6518.365 & 2.830 & -2.465 & 56.0& 48.5&     &     & 30.4&     & 27.3& 57.6& 69.4\\
6574.225 & 0.990 & -5.004 & 26.4&     &	    &	  &	& 38.6&	    & 32.3&     \\
6591.325 & 4.593 & -1.933 &  9.8& 13.9&	    &	  &	& 12.6&	    & 14.9&     \\
6608.024 & 2.279 & -3.879 & 17.3& 16.0&	    &	  &	& 27.8&	    & 24.4&     \\
6625.021 & 1.011 & -5.340 & 15.3& 12.9&	    &	  &	& 27.7&	    & 14.7&     \\
6713.770 & 4.795 & -1.368 & 21.0& 20.3&	    &	  & 12.0& 21.3&	    & 22.5&     \\
6725.353 & 4.103 & -2.107 & 17.3& 19.8&	    &	  &	& 20.1&	    & 23.1&     \\
6733.151 & 4.638 & -1.338 & 26.4& 25.4&	    &	  & 14.1&     &	    & 30.4& 22.3\\
6739.520 & 1.560 & -4.773 & 12.0&     &	    &	  &	& 18.1&	    &	  &     \\
6750.150 & 2.424 & -2.549 & 69.2& 71.4& 41.6&	  &	&112.0&	    & 81.0& 98.8\\
6752.705 & 4.640 & -1.130 & 34.1& 35.2& 14.9&	  & 25.2&     & 24.6& 40.5& 44.3\\
6837.016 & 4.593 & -1.644 & 14.9& 15.7&	    &	  &	&     &  7.1& 16.4& 19.2\\
6857.243 & 4.076 & -2.040 & 21.2& 23.8& 11.4&	  &	& 33.2&	    & 22.9& 20.8\\
\hline
\multicolumn{3}{c}{Fe\,{\sc ii}}\\
\hline
5425.257 & 3.200 & -3.177 & 45.5&     & 53.2& 54.8&     & 38.5& 63.6&     & 45.3\\
5991.376 & 3.153 & -3.563 & 30.3& 36.7& 38.0& 37.0&	& 22.6& 43.4& 31.4& 26.9\\
6084.111 & 3.200 & -3.768 & 20.9& 19.5& 24.9& 38.0&	&     & 28.9& 20.9& 18.0\\
6149.258 & 3.889 & -2.630 & 35.9& 40.9& 46.8& 49.4& 45.7&     & 53.4& 37.9&     \\ 
6247.557 & 3.892 & -2.281 & 52.0& 60.7& 69.3& 79.7& 77.0& 47.9& 77.8& 58.7& 52.9\\
6369.462 & 2.891 & -4.072 & 18.1& 23.9&	    &	  & 23.1& 13.2& 23.1&	  &     \\
6432.680 & 2.891 & -3.520 & 39.3& 49.1& 55.6&	  & 49.4& 35.7& 67.0& 41.2&     \\
6456.383 & 3.903 & -2.115 & 46.2& 73.3& 77.1& 87.7& 79.7&     & 91.6& 68.9& 66.1\\
7222.394 & 3.889 & -3.213 & 20.0&     & 24.3&	  &	&     & 34.6&	  & 19.2\\
7449.335 & 3.889 & -3.176 & 20.0& 24.9& 28.1& 27.1&	& 23.9&     & 28.1& 22.0\\
7515.831 & 3.903 & -3.432 & 10.6&     & 16.3&	  & 22.3&     &     &	  &     \\
\hline
\multicolumn{3}{c}{Ni\,{\sc i}}\\
\hline
5082.339 & 3.658 & -0.489 & 72.4&     & 50.8&     & 42.8& 73.0& 50.5&     &     \\
5094.423 & 3.833 & -1.024 & 33.9& 29.0&     &     & 11.6& 36.2& 15.4& 30.1&     \\
5115.389 & 3.834 & -0.119 & 83.3& 77.5& 53.6& 54.4& 56.1&101.1& 66.0& 83.4& 86.1\\
6111.066 & 4.088 & -0.768 & 38.0&     & 12.7& 21.4& 16.1&     & 16.8& 29.4& 30.4\\
6130.130 & 4.266 & -0.906 & 23.0& 20.4&  8.3&     &  7.9& 22.1&     & 22.8&     \\
6175.360 & 4.089 & -0.491 & 52.8& 50.2& 23.1& 26.9& 27.6& 51.2& 29.4& 46.7& 49.1\\
6176.807 & 4.088 & -0.235 & 67.9& 59.4& 34.5& 37.7& 37.9& 64.9& 41.1& 62.7& 63.8\\
6204.600 & 4.088 & -1.060 & 23.6& 21.8&     &     & 10.0& 25.0& 15.4& 19.0&     \\
6378.247 & 4.154 & -0.811 & 32.4& 27.9& 12.0& 49.8& 10.2& 29.9& 17.5& 29.8&     \\
6643.629 & 1.676 & -2.022 & 96.7& 84.7& 41.6& 31.7& 47.3&103.0& 53.4&     &114.7\\
6772.313 & 3.658 & -0.947 & 52.4& 46.1& 22.9&     & 26.7& 53.5& 31.7& 43.4& 48.5\\
\hline
\end{tabular}

\end{table*}

\subsection{Oxygen}

Oxygen abundances were estimated using the O\,{\sc i} lines at
 7771-775\AA. Each line was synthesised independently to achieve the
 best fits to our spectra.  A differential LTE [O/H] abundance was
 calculated for each line in each star.  Barklem et al. (2000) van der
 Waals parameters were used and $\log gf$ values were taken from
 R03. These O\,{\sc i} lines are formed high in the atmosphere and are
 subject to NLTE effects. It would have been better to try and
 determine oxygen abundances using the weak [O\,{\sc i}]~6300\AA\ line,
 but our data were of insufficient quality to attempt this. 
 Instead we have made a (temperature and gravity dependent) NLTE correction of
 order 0.1-0.2\,dex to the oxygen abundances using the relationship
 derived for field dwarfs by R03. The NLTE corrections proposed by
 Gratton et al. (1999) are similar to a few hundredths of a
 dex. Further comments on the oxygen abundances derived in this manner
 are deferred to section~\ref{meanabundance}.

\subsection{Lithium} 

LTE Li abundances were determined using the Li\,{\sc i} resonance lines at
6707.7\AA\ and 6707.9\AA.  Barklem et al. (2000) van der Waals
parameters were used and $\log gf$ values were taken from the NIST
database -- listed as quality 'A'.  The Li feature was synthesised as a
doublet; the nearby Fe\,{\sc i} line at 6707.4\AA\ was included. No
solar measurement was possible due to the inherent weakness of this
line in the Sun's Li-depleted photosphere. NLTE corrections were
applied using the code provided by Carlsson et al. (1994).

\subsection{Carbon, Silicon, Sulphur, Magnesium, Calcium, Titanium and Nickel}

Lines for neutral species of these elements were taken from the lists
of R03, E95 and Schuler et al. (2003). We chose
lines which were likely to be unblended or features where lines of the
same species were blended together. Accurate $\log gf$ values were
generally not available, so astrophysical $\log gf$ values were
obtained by an inverse solar analysis, assuming solar abundances of
$A$(C)$=8.51$, $A$(Si)$=7.62$, $A$(S)$=7.34$, $A$(Mg)$=7.54$,
$A$(Ca)$=6.33$, $A$(Ti)$=4.90$ and $A$(Ni)$=6.23$, which were derived by
R03. van der Waals parameters were taken from Barklem et al. (2000)
where available.  The final quoted abundances are differential with
respect to the assumed solar values.

\section{Results}

\subsection{Chemical Abundances}
\label{abundances}

Table~\ref{table-results} contains the mean abundances for each
Blanco~1 target. All results (apart from Li) are differential with
respect to the Sun.  The quoted uncertainties for each star in this
Table are the standard error in the mean abundance from all the
features used for that star and element; they do not include the
effects of uncertainties in the atmospheric parameters (see
section~\ref{errors}).

The RVs determined for our targets in Table~1 are reasonably
consistent with each other and all lie within 2$\sigma$ of the mean
cluster value deduced by E95 (3.9$\pm$0.7 km\,s$^{-1}$). Three of our
targets (W8, W60, W63) have RV measurements listed in E95
which are reasonably consistent with those reported here. Four targets
(W64, W113, ZS58, ZS141) have RVs and $v \sin i$ listed by JJ99 and
again they agree within their errors with those found here.

It is worth mentioning that both E95 and Westerlund et al. (1988)
doubted the membership credentials of W8. Its distance modulus, derived
from Stromgren photometry, is about 0.3\,mag lower than most cluster
members. The inclusion or not of W8 has no significant
effect on our final results for the abundances of Blanco~1.  However,
we are quite confident on the basis of their RVs that all of these
objects are members of Blanco 1 and are either single, long period
binary systems or pole-on binary systems.  None of the spectra show any
signs of contamination from a secondary star.

\begin{table*}
\caption{Abundances for target stars. Errors quoted are the
  standard errors in the measured abundances 
  ($\sigma / \surd n$). Atmospheric uncertainties
  (detailed in Table~\ref{table-atmuncerts}) should also be
  considered. Columns labelled ``n'' identify the number of features used
  when obtaining each value. Solar abundances were determined for iron
  and oxygen, due to the availability of accurate laboratory gfs. In
  all other cases the solar abundance was fixed at the values listed in
  the last row. Abundances are quoted differentially with
  respect to the Sun apart from those for Li. LTE abundances are given
  and NLTE abundances are also listed for Li and O as discussed in the text.}
\label{table-results}
\begin{tabular}{lcccccccccccc}
\hline
ID      & $A$(Li) & $A$(Li)  & n  & [C/H]  & n
& [O/H] & [O/H] & n  & [Mg/H] & n  & [Si/H] & n \\ 
        & LTE     & NLTE     &    &        &
& LTE   & NLTE  &    &        &    &        & \\
\hline
W64     &  2.69$\pm$0.05& 2.65 & 1  & $-0.13\pm0.08$ & 3  & $+0.03\pm0.06$&$-0.09$  & 3  & $-0.17\pm0.04$  & 4  & $-0.14\pm0.02$   & 7 \\
ZS58    &  3.31$\pm$0.06& 3.09 & 1  & $-0.05\pm0.10$ & 1  & $+0.05\pm0.05$&$-0.10$  & 2  & $-0.07\pm0.08$  & 4  & $-0.16\pm0.04$   & 6 \\
ZS141   &  2.89$\pm$0.05& 2.81 & 1  & $-0.16\pm0.17$ & 3  & $+0.02\pm0.06$&$-0.11$  & 3  & $+0.01\pm0.05$  & 3  & $-0.06\pm0.02$   & 7 \\
W113    &  3.14$\pm$0.05& 3.00 & 1  & $-0.17\pm0.01$ & 2  & $+0.01\pm0.02$&$-0.12$  & 3  & $-0.07\pm0.02$  & 4  & $-0.04\pm0.02$   & 5 \\    
W60     &  3.23$\pm$0.12& 3.11 & 1  & $-0.01\pm0.06$ & 3  & $+0.40\pm0.06$&$+0.24$  & 3  & $-0.10\pm0.05$  & 3  & $+0.05\pm0.05$   & 2 \\
W63     &  3.13$\pm$0.07& 3.03 & 1  & $-0.15\pm0.01$ & 3  & $+0.18\pm0.02$&$+0.01$  & 3  & $-0.14\pm0.02$  & 4  & $-0.05\pm0.04$   & 6 \\
W8      &  2.60$\pm$0.13& 2.54 & 1  & $+0.13\pm0.02$ & 3  & $+0.35\pm0.05$&$+0.19$  & 3  & $-0.06\pm0.05$  & 3  & $+0.03\pm0.04$   & 7 \\
W38     &  3.06$\pm$0.10& 2.97 & 1  & $+0.00\pm0.02$ & 3  & $+0.29\pm0.01$&$+0.10$  & 3  & $-0.14\pm0.02$  & 4  & $-0.07\pm0.04$   & 7 \\ 
$A$(X)$_{\odot}$   &   &               &   &  8.51 & 3  & $8.89\pm0.01$       &  & 3  &  7.54     & 4  & 7.62            & 7 \\ 
\hline
        &                 &    &                &    &                 &    &                 &    &                  &   \\
\end{tabular}
\begin{tabular}{lcccccccccc}
\hline
ID      & [S/H]           & n  & [Ca/H]          & n  & [Ti/H]
& n  & [Fe/H] & n &[Ni/H]             & n  \\
\hline
W64     & $+0.04\pm0.09$  & 1  &  $-0.17\pm0.02$ & 3  & $-0.16\pm0.03$ & 4  & $-0.04\pm0.02$ & 33 & $-0.20\pm0.03$ & 10 \\	      
ZS58    & $+0.24\pm0.14$  & 2  &  $-0.05\pm0.04$ & 3  & $+0.03\pm0.03$ & 5  & $-0.05\pm0.02$ & 24 & $-0.26\pm0.01$ & 6  \\       
ZS141   & $+0.00\pm0.08$  & 1  &  $-0.05\pm0.04$ & 3  & $+0.00\pm0.05$ & 4  & $+0.09\pm0.02$ & 34 & $-0.10\pm0.02$ & 9  \\	      
W113    & $-0.12\pm0.06$  & 2  &  $+0.04\pm0.04$ & 3  & $-0.04\pm0.05$ & 5  & $+0.09\pm0.02$ & 35 & $-0.03\pm0.01$ & 9  \\	      
W60     & $+0.02\pm0.02$  & 2  &  $-0.15\pm0.05$ & 2  & $-0.18\pm0.12$ & 1  & $+0.11\pm0.03$ & 13 & $-0.15\pm0.05$ & 6  \\	      
W63     & $-0.02\pm0.01$  & 2  &  $-0.02\pm0.03$ & 3  & $-0.12\pm0.03$ & 3  & $+0.05\pm0.02$ & 24 & $-0.17\pm0.03$ & 11  \\	      
W8      & $+0.10\pm0.04$  & 2  &  $+0.02\pm0.03$ & 3  & $+0.07\pm0.06$ & 3  & $+0.09\pm0.02$ & 25 & $-0.03\pm0.03$ & 10  \\	      
W38     & $-0.07\pm0.01$  & 2  &  $-0.11\pm0.03$ & 3  & $-0.01\pm0.10$ & 2  & $-0.03\pm0.02$ & 24 & $-0.17\pm0.04$ & 9  \\	      
$A$(X)$_{\odot}$     & 7.34   & 2  &  6.33 & 3  &  4.90 & 5  & $7.44\pm 0.01$ &  43 & 6.23     & 11  \\ 
\hline
\end{tabular}

\end{table*}

\begin{table*}
\caption{Abundance uncertainties due to estimated atmospheric
  uncertainties.The quadratic sum of uncertainties due to effective
  temperature ($\pm$100~K), $\log g$ ($\pm$0.2), [M/H] ($\pm$0.1 dex)
  and microturbulence ($\pm$0.2 km\,s$^{-1}$) are presented for each
  star. The first number is the net uncertainty in [X/H], the number in
  brackets is the net uncertainty in [X/Fe].} 
\label{table-atmuncerts}
\begin{tabular}{llllllllll}
\hline 
      & W64       & ZS58      & ZS141     & W113       & W60       & W63          & W8         & W38           \\
\hline
Fe    & 0.07      & 0.06      & 0.06      & 0.07       & 0.06      & 0.05         & 0.05       & 0.05          \\
Li    & 0.11(0.07)& 0.11(0.07)& 0.10(0.06)& 0.09(0.05) & 0.08(0.06)& 0.07(0.04)   & 0.07(0.04) & 0.07(0.05)    \\
C     & 0.10(0.14)& 0.09(0.12)& 0.09(0.13)& 0.09(0.13) & 0.08(0.09)& 0.07(0.10)   & 0.07(0.10) & 0.07(0.09)    \\
O     & 0.12(0.16)& 0.11(0.14)& 0.14(0.18)& 0.09(0.14) & 0.06(0.09)& 0.06(0.10)   & 0.06(0.09) & 0.06(0.09)    \\
Si    & 0.03(0.07)& 0.02(0.05)& 0.02(0.05)& 0.02(0.05) & 0.03(0.05)& 0.04(0.03)   & 0.04(0.04) & 0.03(0.04)    \\
S     & 0.10(0.14)& 0.09(0.12)& 0.08(0.12)& 0.07(0.12) & 0.05(0.08)& 0.05(0.09)   & 0.05(0.08) & 0.05(0.07)    \\
Mg    & 0.06(0.06)& 0.06(0.06)& 0.07(0.07)& 0.06(0.05) & 0.06(0.06)& 0.05(0.04)   & 0.06(0.05) & 0.05(0.05)    \\
Ca    & 0.10(0.07)& 0.09(0.06)& 0.09(0.05)& 0.09(0.05) & 0.06(0.05)& 0.06(0.03)   & 0.06(0.04) & 0.06(0.04)    \\
Ti    & 0.11(0.07)& 0.10(0.07)& 0.10(0.05)& 0.10(0.05) & 0.08(0.06)& 0.07(0.04)   & 0.07(0.04) & 0.07(0.05)    \\
Ni    & 0.05(0.02)& 0.06(0.02)& 0.06(0.02)& 0.08(0.06) & 0.06(0.04)& 0.06(0.02)   & 0.06(0.03) & 0.06(0.04)    \\
\hline
\end{tabular}

\end{table*}

\subsection{Uncertainties}

\label{errors}

\subsubsection{EW and $\log gf$ errors}

The errors quoted in Table~\ref{table-results} encapsulate
uncertainties in the individual EW measurements and the $\log gf$
values for the lines.  These were generally derived from the standard
deviation of differental abundances for each line. In cases where only
one line was available we used $\delta$EW = $(rp)^{0.5}/{\rm SNR}$
(where $r$ is the physical width of the line, and $p$ is the pixel
size, both in \AA) and translated this into an abundance uncertainty.
For lithium, the atomic parameters are well understood and introduce
additional abundance uncertainties of only $\simeq$0.01 dex.

The scatter in the abundances listed in
Table~\ref{table-results} is much larger than the individual error
bars. However, we have yet to include a contribution from the uncertain
atmospheric parameters for each star.

\subsubsection{Atmospheric parameters}

$T_{\rm{eff}}$ uncertainties are the most important contributor
to uncertainties in abundance for all elements except C and S (see below). 
To estimate a realistic $T_{\rm eff}$
uncertainty we compare $T_{\rm eff}$ derived from the spectroscopy with
that derived using the relationship between $T_{\rm eff}$ and $V-K$
colour index from Alonso, Arribas \& Martinez-Rogers (1996). 
A small correction was made to
the 2MASS $K$ photometry (using formulae in Carpenter 2001) in order to
convert it to the Carlos Sanchez Telescope (TCS) system 
used by Alonso et al. A reddening $E(V-K)=0.055$ is assumed,
corresponding to $E(B-V)=0.02$ for the cluster determined by Westerlund et
al. (1988). We chose this colour index
because (i) the data is available for all our targets; (ii) it is very
sensitive to $T_{\rm eff}$; (iii) it is almost independent of the
photospheric composition and gravity and (iv) is unlikely to be significantly
affected by chromospheric activity (Stauffer et al. 2003). This latter
point could be important in a young cluster like Blanco~1 where
chromospheric activity can lead to blue excesses and possible problems when
using $B-V$ or Stromgren photometry to determine $T_{\rm eff}$ or $\log g$.

$T_{\rm eff}$ values 
derived from the $V-K$ photometry and from the spectroscopy are listed
in Table~\ref{table-params}. A comparison of the two
$T_{\rm eff}$ determination methods yields a mean difference
($T_{V-K}-T_{\rm spec}$) of $-13\pm36$\,K with a standard deviation of 102\,K.
As the precision of the photometry leads to uncertainties of only 
$\simeq 50$\,K in $T_{V-K}$ then most of this scatter must be due to
$T_{\rm spec}$ uncertainties of $\simeq 100$\,K. The agreement between the
two scales lends some confidence that there are no major
systematic uncertainties in the temperatures we have used and no
problems with our ionization balance temperatures caused by possible NLTE
overionization effects in the Fe\,{\sc ii} lines -- which may become more apparent in cooler
stars ($<5500$\,K -- see Schuler et al. 2003; Allende-Prieto et
al. 2004). In addition we have checked plots of abundance from the
Fe\,{\sc i} lines versus lower excitation potential and none show any
significant trends that would indicate a $T_{\rm eff}$ error of more
than $\pm 150$\,K.

Having used a cluster isochrone in the $T_{\rm eff}-\log g$ plane to
determine $\log g$, then an uncertainty in $T_{\rm eff}$ naturally
leads to an uncertainty in $\log g$. In fact this uncertainty is small,
but we choose (conservatively) to allow $\log g$ to vary by $\pm 0.2$.
Because the isochrone $\log g$ varies very slowly with $T_{\rm eff}$,
the uncertainties in $T_{\rm eff}$ and the 
assumed $\log g$ errors are essentially
uncorrelated. The $\log g$ uncertainty is dominant for the C and S
abundance determinations, but less important than the $T_{\rm eff}$
uncertainties for all the other elements. We adopt conservative
microturbulence uncertainties of $\pm 0.2$\,km\,s$^{-1}$ and
atmospheric metallicity uncertainties of $\pm0.1$\,dex. These
contribute 0.02-0.03\,dex abundance uncertainty to the O, C and Fe
abundances but add a negligible amount to the overall error budget for
the other elements.

Table~\ref{table-atmuncerts} details the quadratic sum of uncertainties
in [X/H] due to the atmospheric parameter uncertainties, making the
assumption that the different sources of error are independent. These
have been estimated by repeating the abundance analysis for each
star/element after perturbing their atmospheric parameters.  We have
also estimated total uncertainties on [X/Fe]. These are smaller or
larger than corresponding uncertainties in [X/H] depending on whether
changes in the atmospheric parameters cause changes in the derived
abundances which are in the same (e.g. Ca, Ni) or contrary (e.g. C, O,
S) direction to those in the Fe abundance.  The errors in
Table~\ref{table-atmuncerts} should be combined in quadrature with
those quoted in Table~\ref{table-results} in order to obtain the
overall (internal) errors on the abundances for each star.

\subsection{Mean cluster abundances}
\label{meanabundance}

\begin{table*}
\caption{Weighted mean abundances for Blanco 1, based on observations
  in this work. Errors are the (weighted) standard error and do not
  include any external uncertainties due to choice of atmospheric model or
  systematic shifts in the $T_{\rm eff}$ scale.}
\label{table-clustermean}
\begin{tabular}{lcccccccc}
\hline
Element  & [X/H]   & $\Delta
T=+100$\,K&  $\chi^{2}_{\nu}$ & $P(>\chi^{2})$ & [X/Fe] &  $\Delta
T=+100$\,K&  $\chi^{2}_{\nu}$ & $P(>\chi^{2})$ \\
\hline
Fe      &  $+0.04\pm0.02$ & $+0.04$ & 1.10 & 0.36 && & & \\
$A$(Li) &  $+2.92\pm0.04$ & $+0.08$ & 2.59 & 0.01 & & & \\
C       &  $-0.04\pm0.04$ & $-0.05$ & 1.68 & 0.11 & $-0.08\pm0.04$&$-0.09$ &0.95 &0.47 \\
O$^{\dag}$&  $+0.06\pm0.11$ &$-0.04$& 2.31 & 0.02 & $+0.02\pm0.11$&$-0.09$ &1.08 &0.38 \\
Mg      &  $-0.11\pm0.02$ & $+0.05$ & 0.60 & 0.76 & $-0.14\pm0.02$&$+0.00$ &0.55 &0.80 \\
Si      &  $-0.06\pm0.02$ & $+0.02$ & 2.44 & 0.02 & $-0.09\pm0.02$&$-0.03$ &0.41 &0.89 \\
S       &  $+0.00\pm0.03$ & $-0.03$ & 1.20 & 0.30 & $-0.04\pm0.04$&$-0.08$ &0.88 &0.52 \\
Ca      &  $-0.06\pm0.03$ & $+0.07$ & 0.79 & 0.60 & $-0.09\pm0.02$&$+0.02$ &1.06 &0.39 \\
Ti      &  $-0.06\pm0.03$ & $+0.09$ & 0.92 & 0.49 & $-0.10\pm0.03$&$+0.05$ &2.16 &0.04 \\
Ni      &  $-0.15\pm0.03$ & $+0.05$ & 1.38 & 0.21 & $-0.18\pm0.01$&$+0.01$ &0.80 &0.59 \\
\hline
\multicolumn{7}{l}{$\dag$ O abundance errors include $\pm0.1$ in NLTE corrections.}
\end{tabular}

\end{table*}

\begin{figure*}
    \centering
    \begin{minipage}[t]{0.45\textwidth}
    \includegraphics[width=71mm]{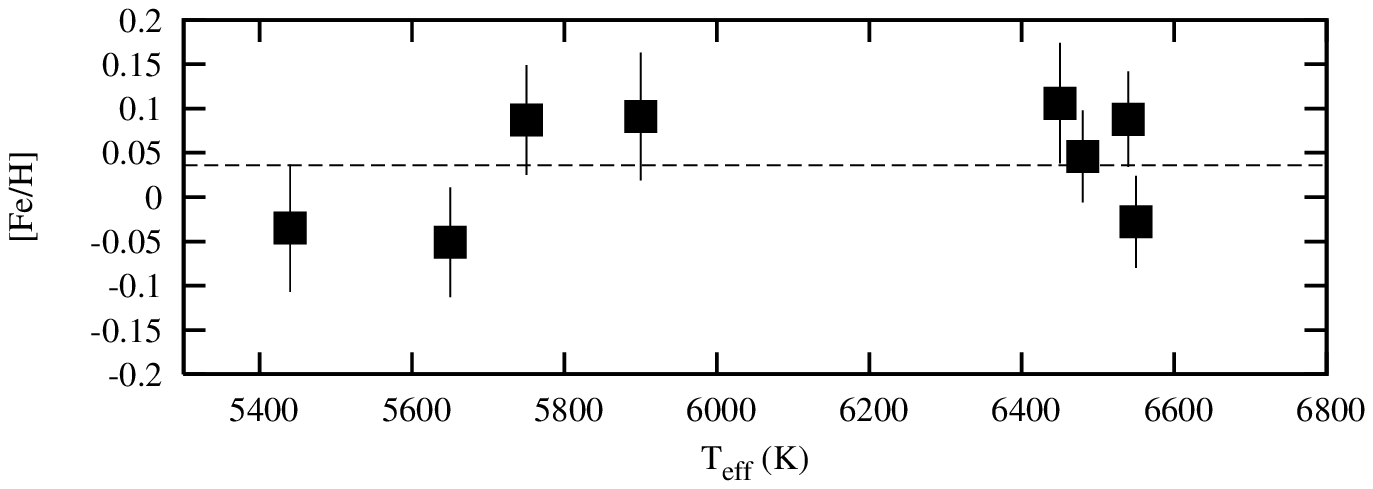}
    \includegraphics[width=71mm]{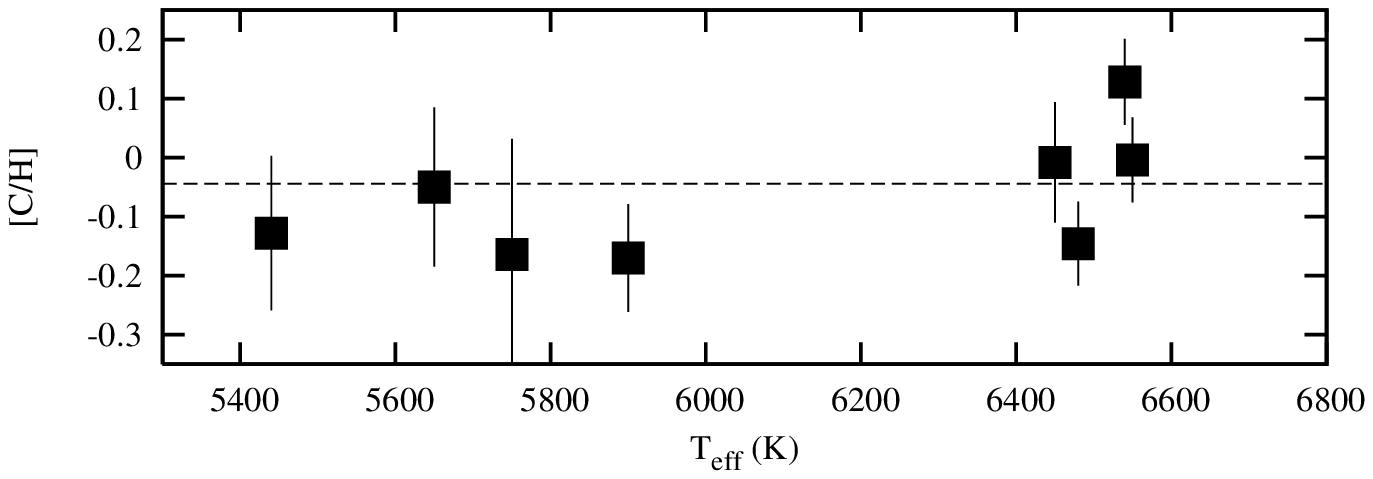}
    \includegraphics[width=71mm]{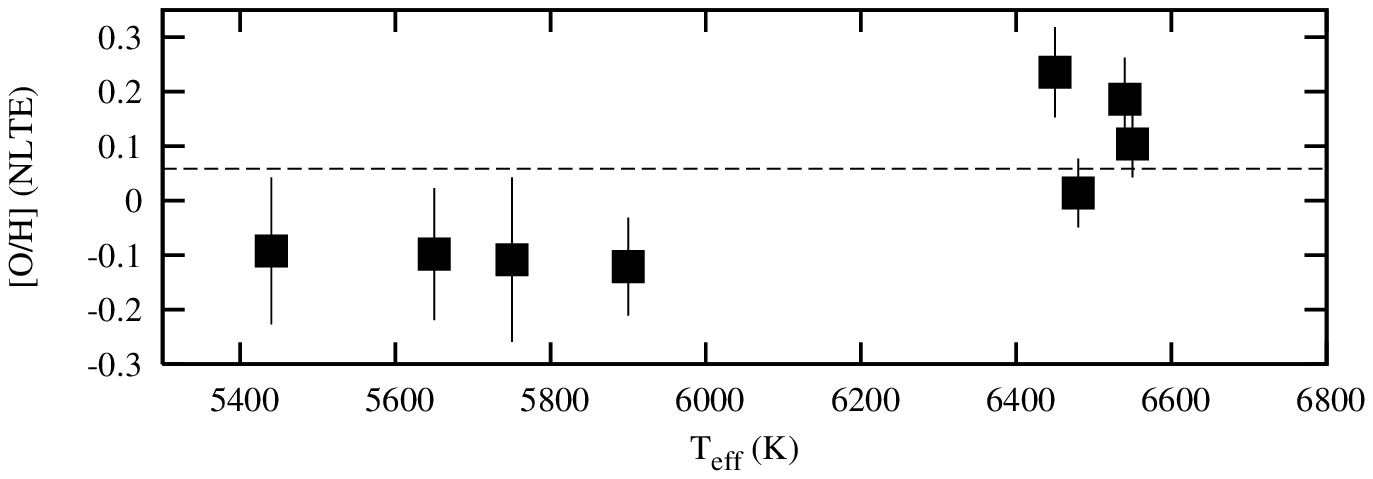}
    \includegraphics[width=71mm]{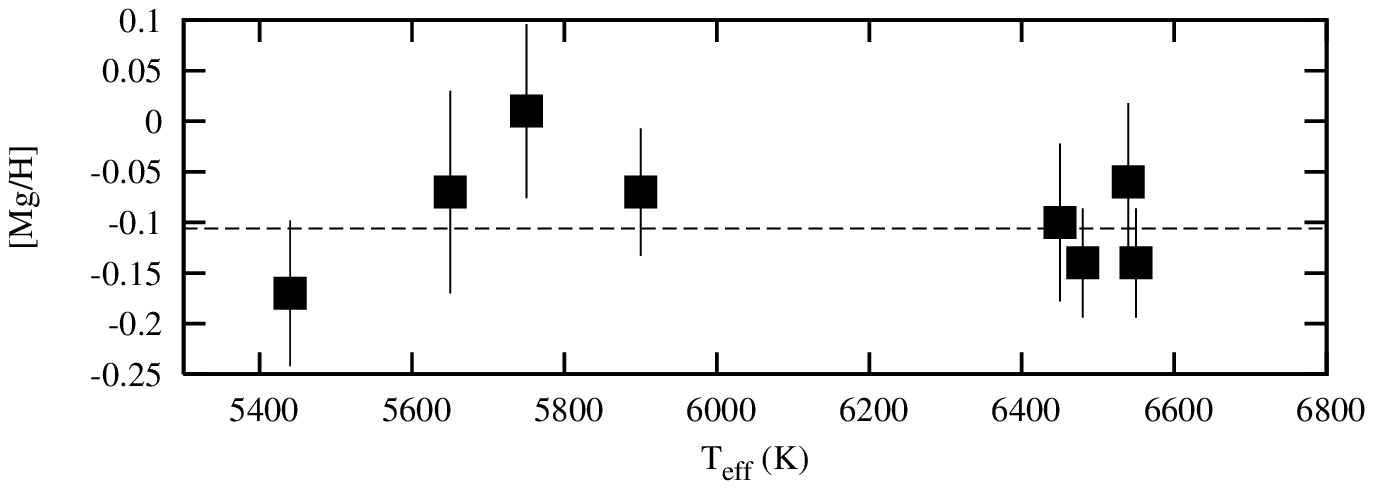}
    \includegraphics[width=71mm]{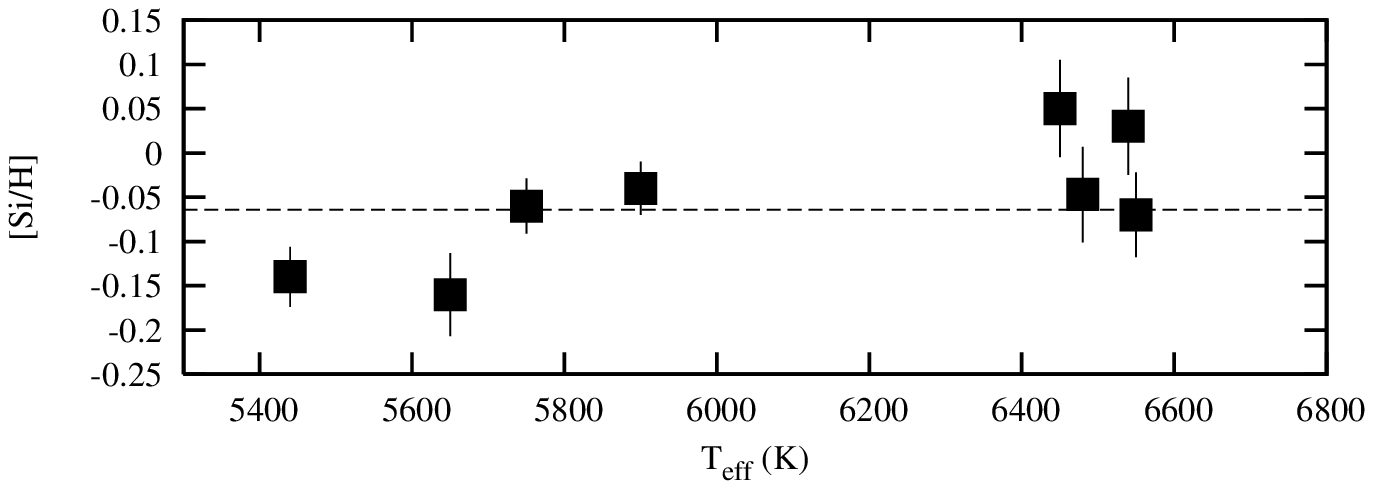}
    \includegraphics[width=71mm]{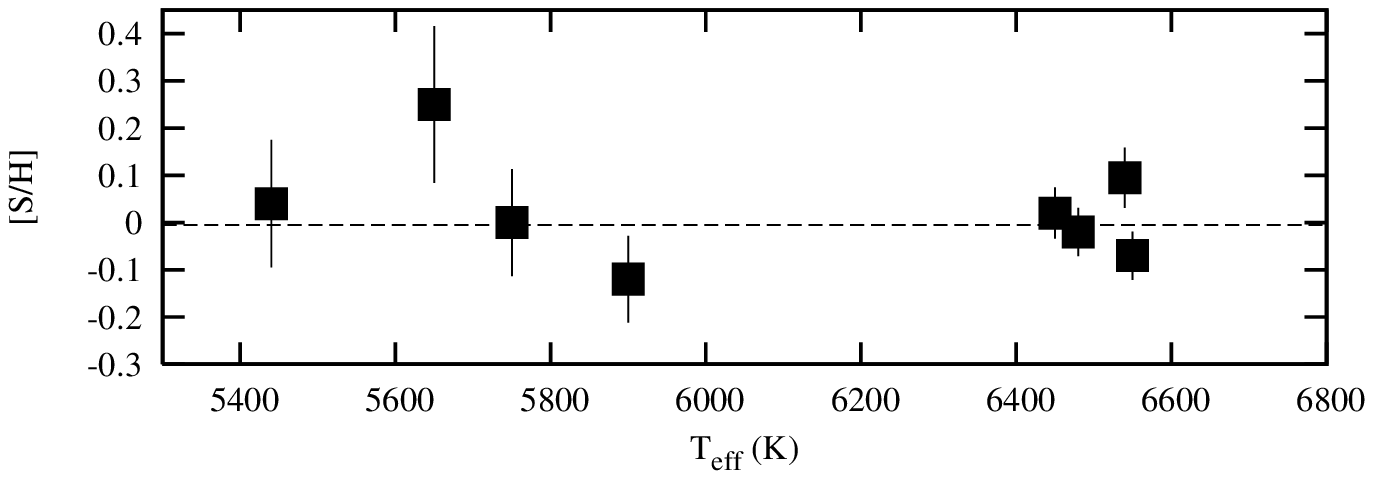}
    \includegraphics[width=71mm]{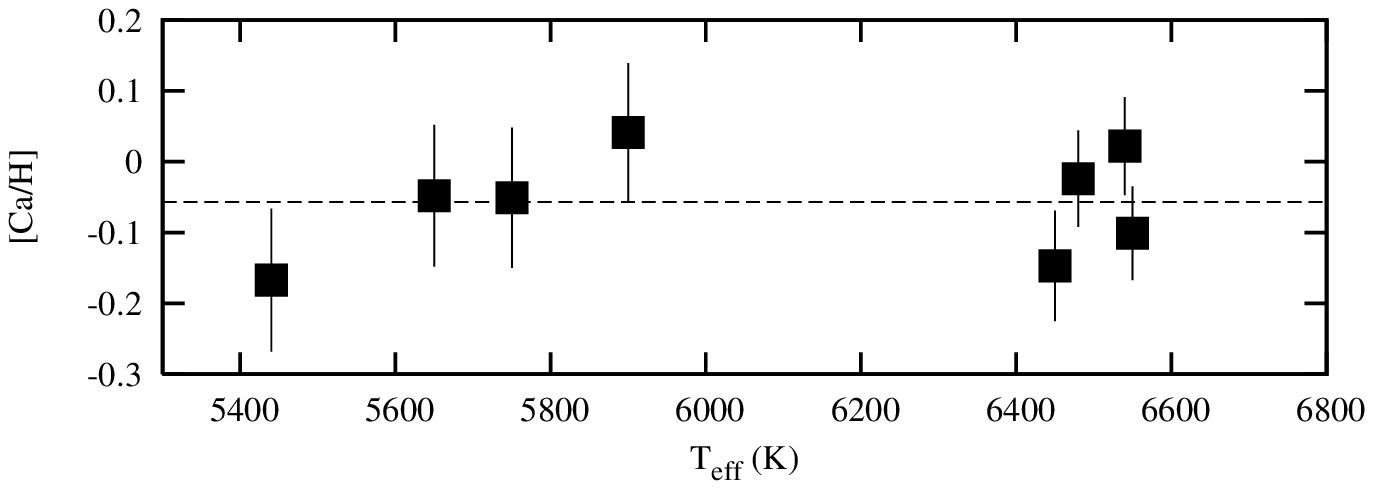}
    \includegraphics[width=71mm]{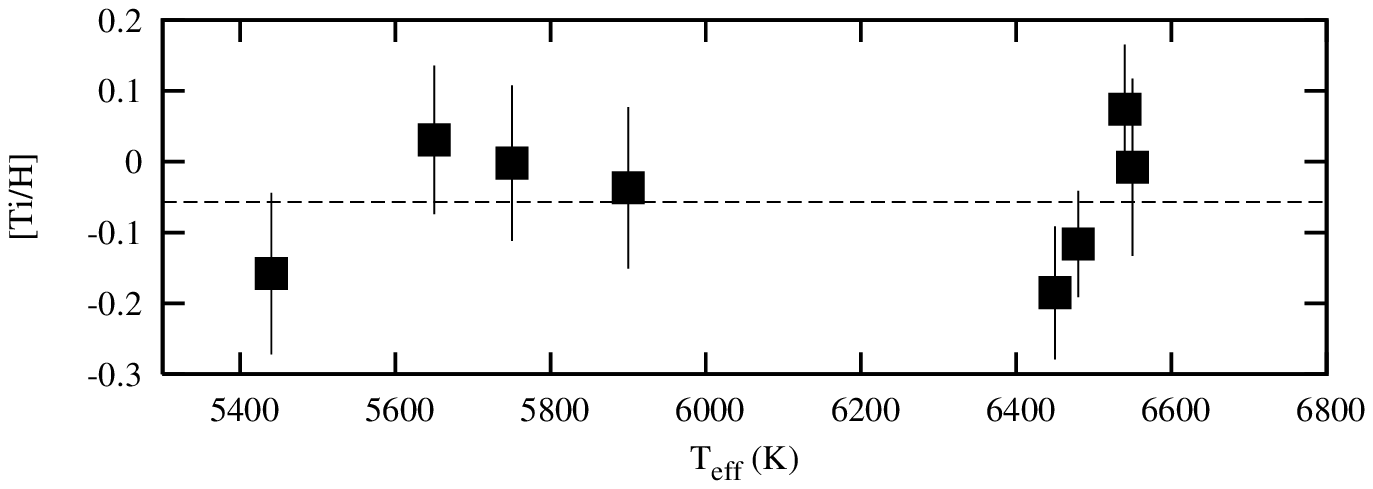}
    \includegraphics[width=71mm]{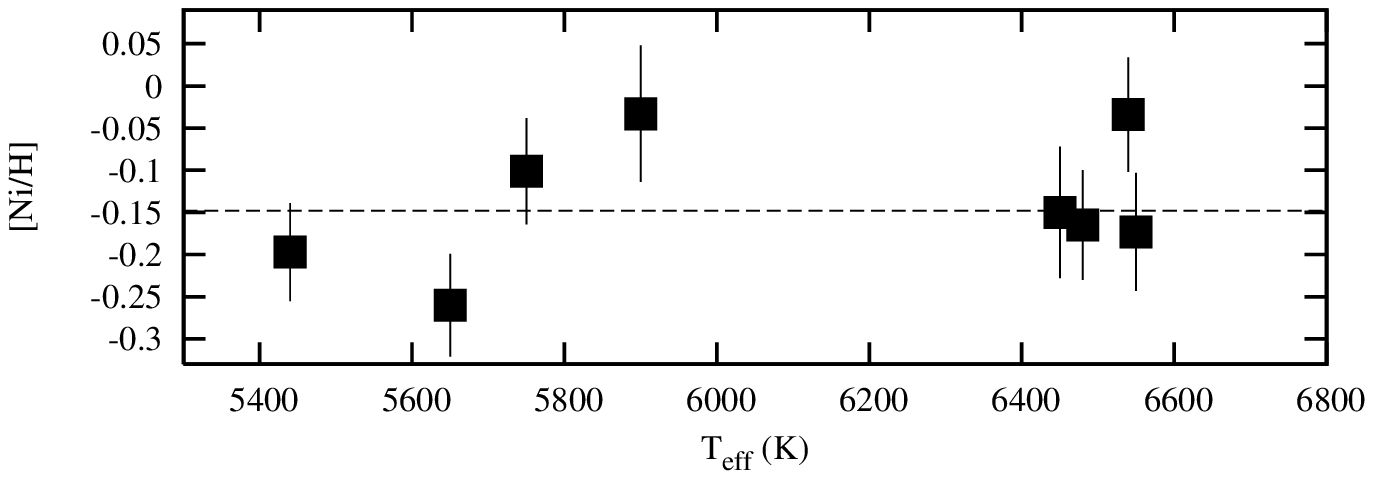}
    \end{minipage}
    \begin{minipage}[t]{0.45\textwidth}
    \includegraphics[width=71mm]{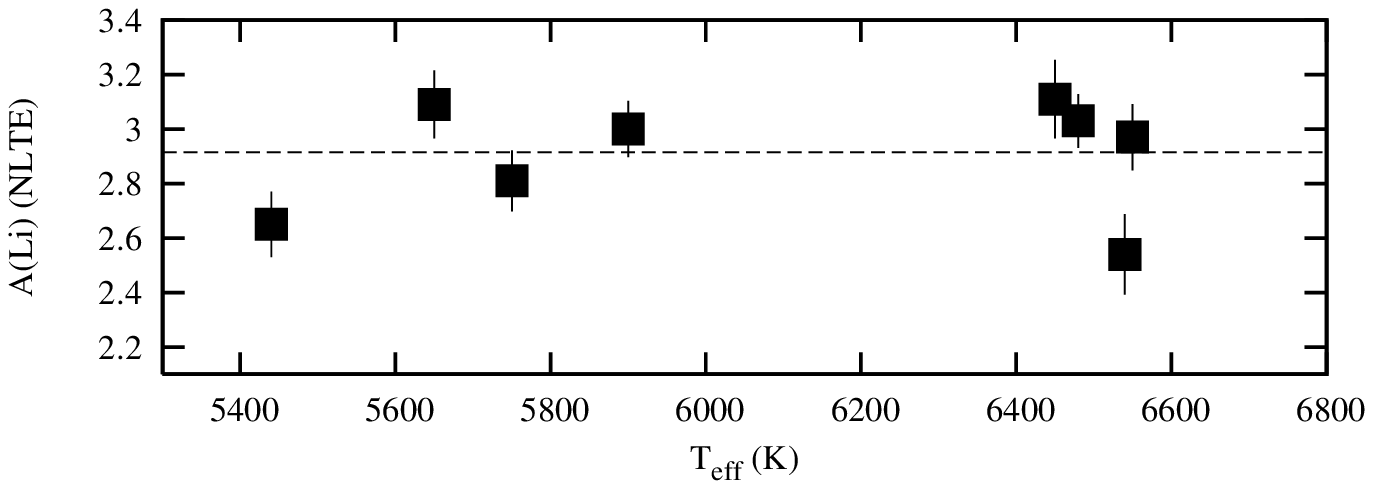}
    \includegraphics[width=71mm]{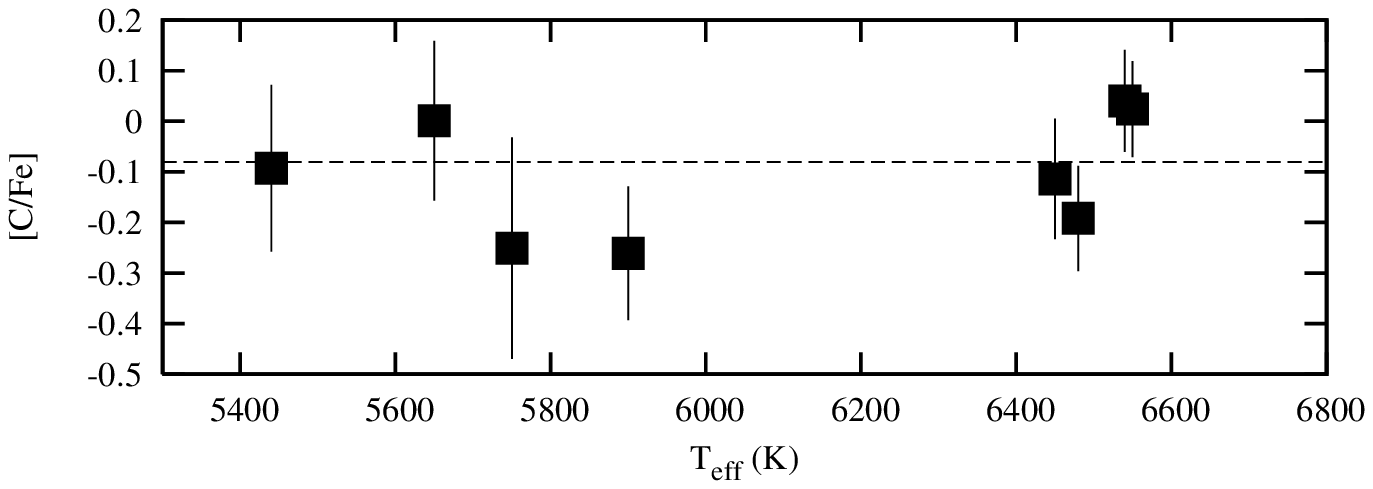}
    \includegraphics[width=71mm]{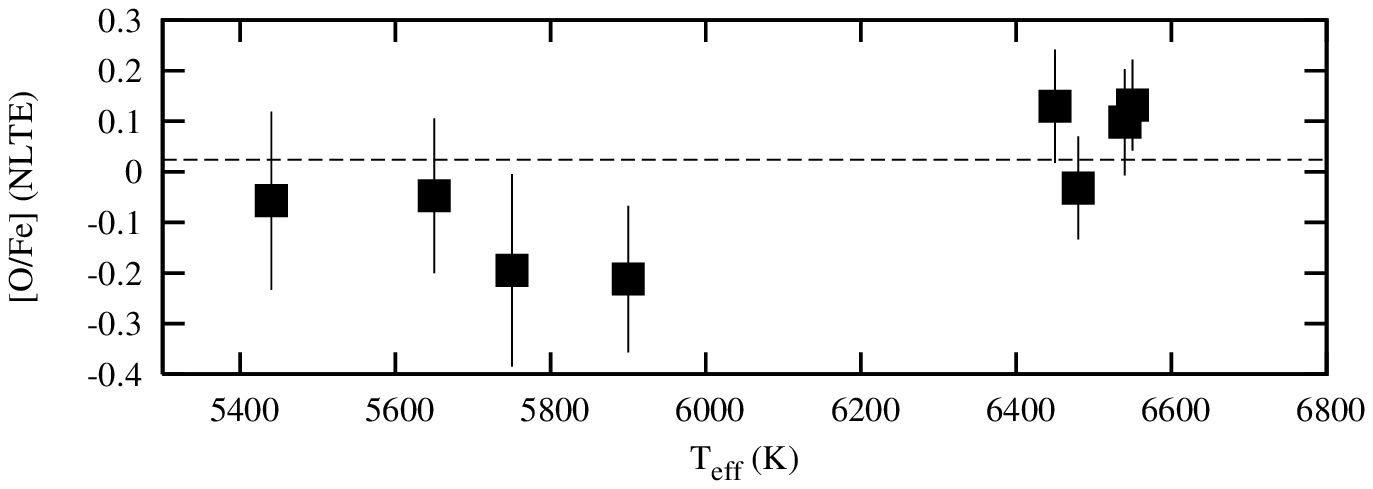}
    \includegraphics[width=71mm]{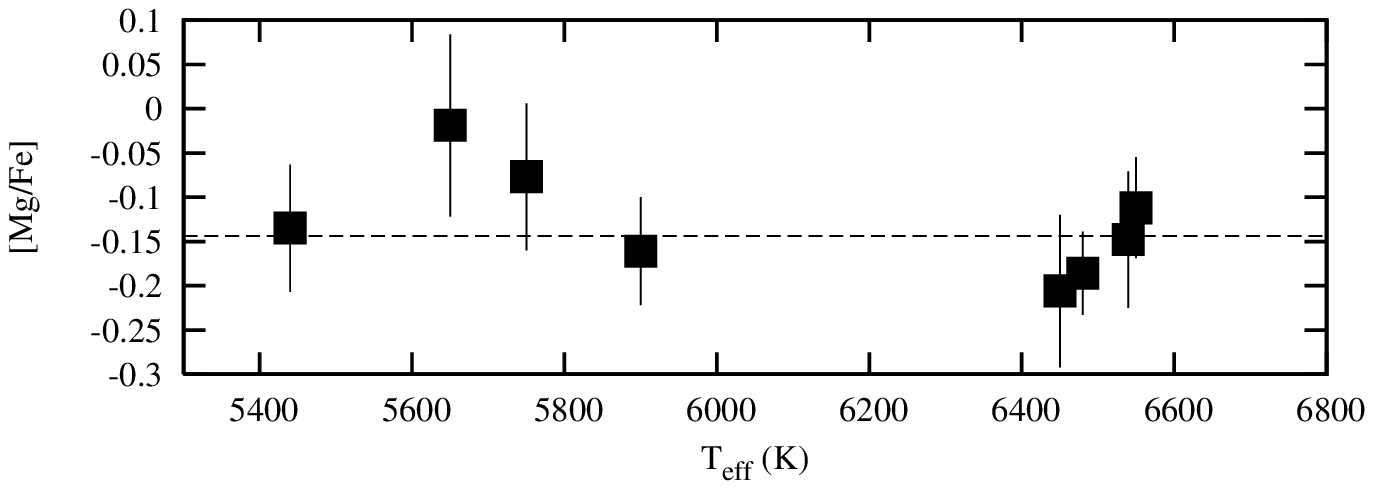}
    \includegraphics[width=71mm]{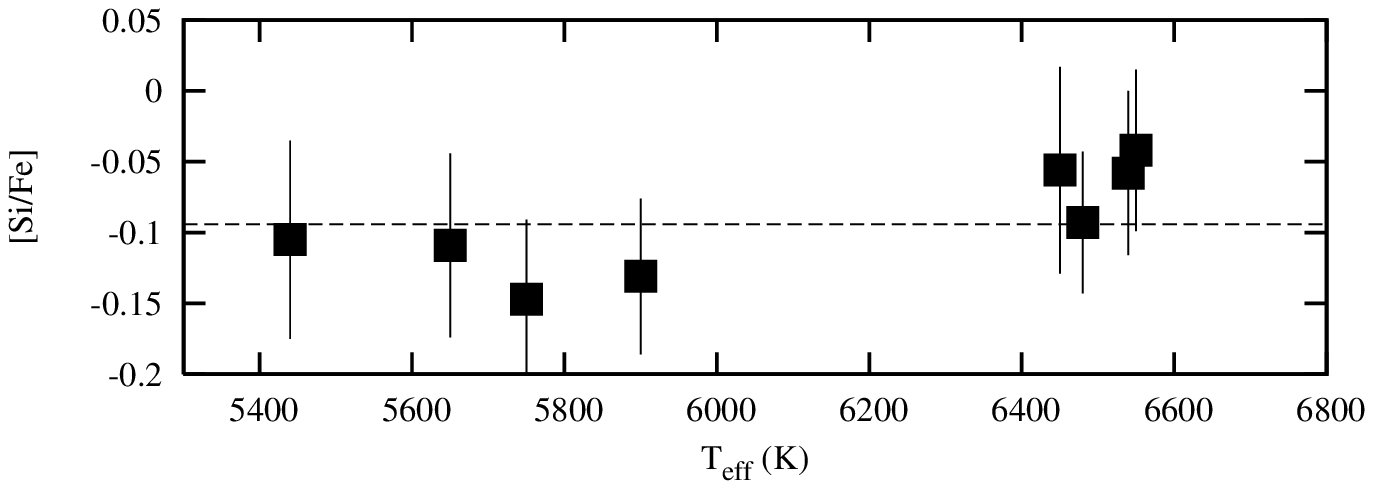}
    \includegraphics[width=71mm]{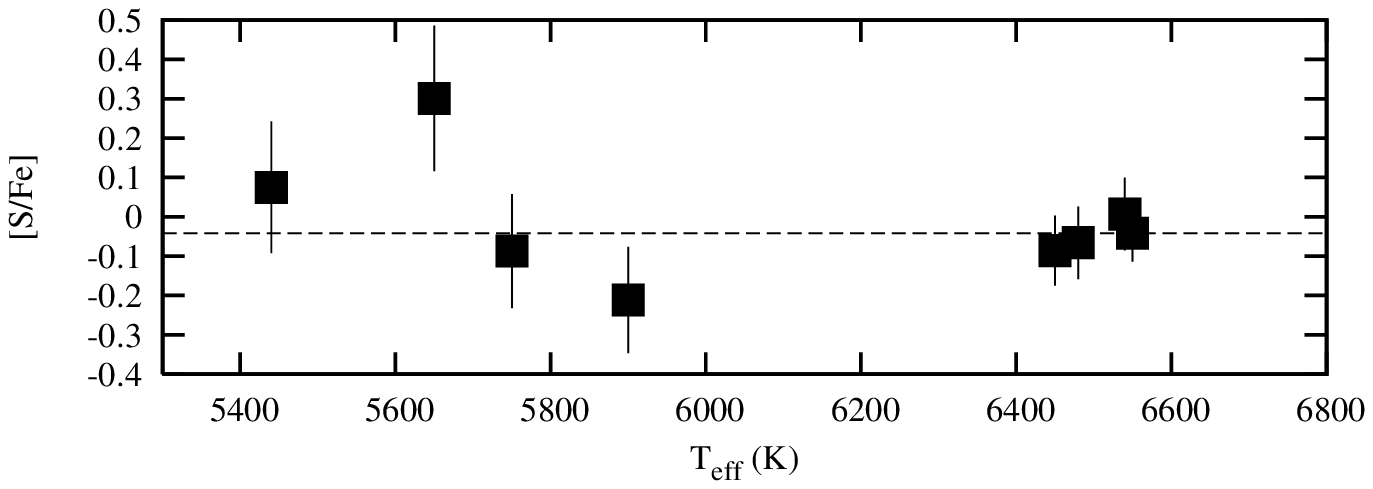}
    \includegraphics[width=71mm]{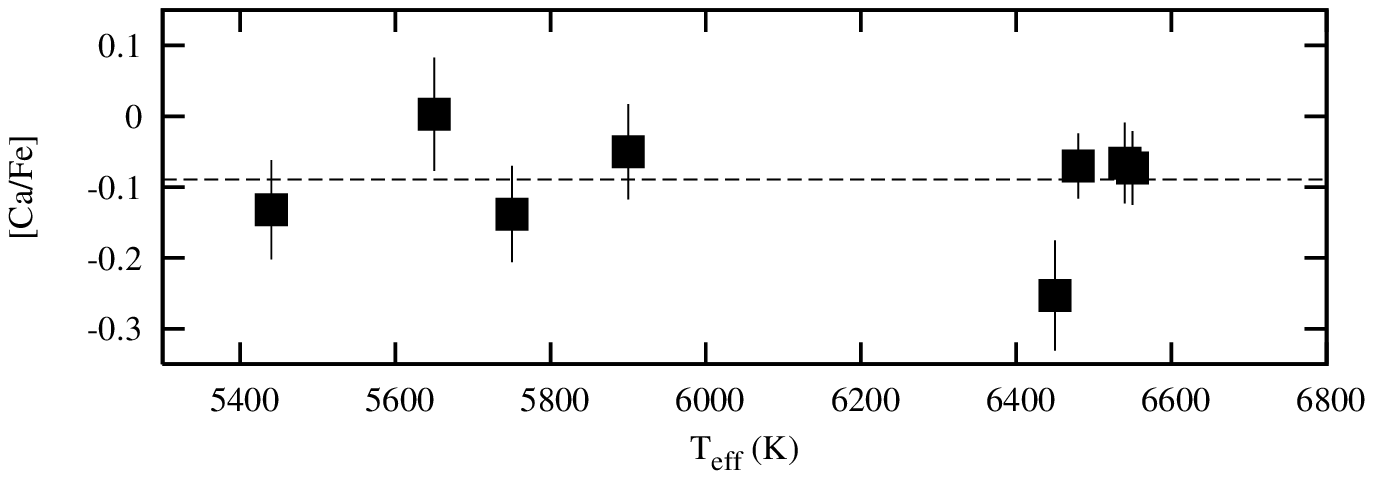}
    \includegraphics[width=71mm]{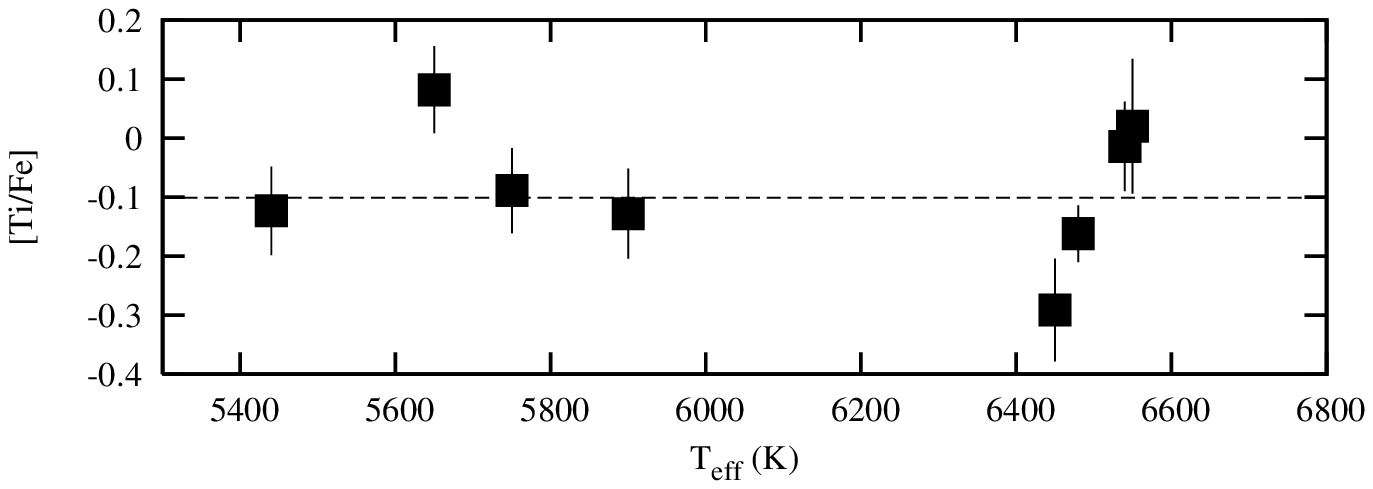}
    \includegraphics[width=71mm]{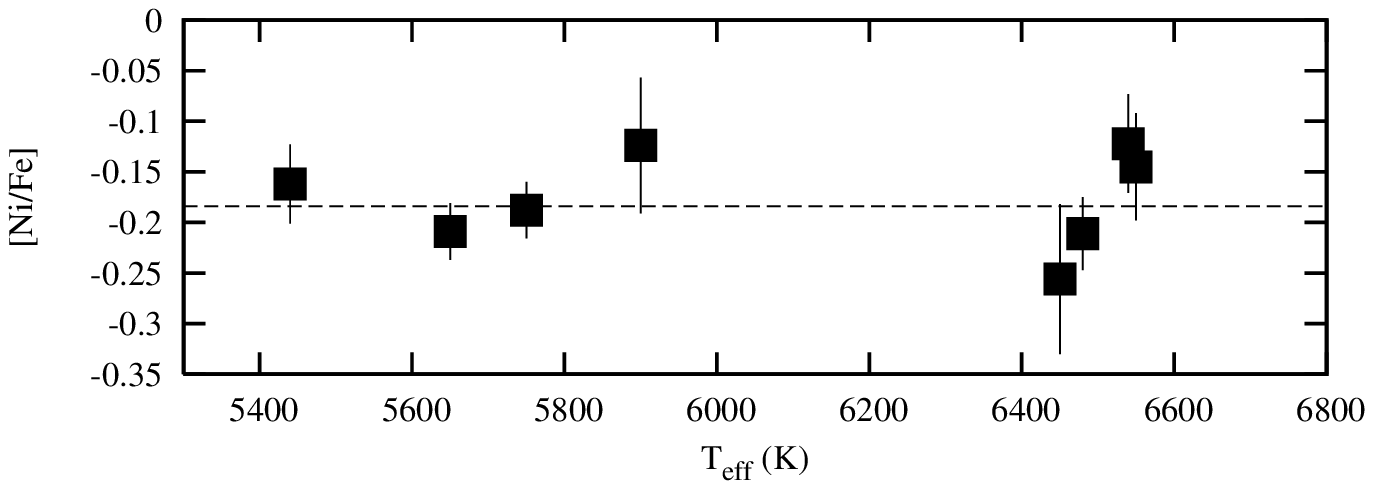}
    \end{minipage}
    \caption{Abundances and abundance ratios versus $T_{\rm
    eff}$. Dashed lines indicate weighted means.}
\label{abunplots}
\end{figure*}

With the internal uncertainties established,
Table~\ref{table-clustermean} lists the weighted mean abundances for
Blanco~1 in the form of [X/H] and [X/Fe], using all eight stars and the
quadratic sum of the uncertainties presented in
Tables~\ref{table-results} and~\ref{table-atmuncerts}. We quote the
standard errors in the weighted mean and also the reduced chi-squared
(for 7 degrees of freedom) of the weighted mean fitted to the data and the
probability that a chi-squared of this size could arise given the
quoted errors. These results are graphically presented in Fig.~\ref{abunplots}.

For Fe, C, Mg, S, Ca, Ti and Ni the scatter in the [X/H] abundance measurements
are consistent with the estimated uncertainties, with
reasonably low reduced chi-squared values. This lends confidence in our methods and
uncertainty estimates for these elements and also suggests that any
star-to-star scatter of abundances within the cluster is smaller than
the uncertainties estimated for each star.  
However, there are three elements 
(Li, O, Si) where a high reduced chi-squared is found. 
This could indicate either
(i) that the elemental abundance genuinely varies from star-to-star,
(ii) that the abundance uncertainties have been underestimated
or (iii) that there is an apparent trend of abundance with $T_{\rm
eff}$ arising from an inadequate treatment of the atmosphere, NLTE
effects or the temperature scale.

For Li it is very likely that explanation (i) applies. Li is known to
be depleted in many cooler ($T_{\rm eff}<5800$\,K) stars among Blanco~1
and the similarly aged Pleiades (e.g. Soderblom et al. 1993; JJ99).
This could account for low Li abundances in ZS141 and W64. Of more
interest is the low Li abundance of W8 with a $T_{\rm eff}\simeq
6500$\,K and a NLTE $A$(Li) which is about 0.5\,dex lower than the
three other stars in the sample with similar $T_{\rm eff}$.  This star
also has a lower Li abundance than any similar stars in the
Pleiades. It is tempting to speculate that this marks the development
of the ``Boesgaard gap'' (Boesgaard \& Tripicco 1986) of Li-depleted
F-stars that is clearly seen in the older (700\,Myr) Hyades
cluster. Steinhauer \& Deliyannis (2004) claim that the gap starts to
form as early as 150\,Myr on the basis of Li-depleted F-stars in
the open cluster M35. 
However, the membership of
W8 in Blanco~1 may still be 
problematic (see section~\ref{abundances}).  If the membership of
W8 can be confirmed and further examples of Li-depleted F-stars were
found in Blanco~1 this would probably indicate that the cluster is a
little older than the Pleiades.

The O abundances of the Blanco~1 stars show a clear trend with $T_{\rm
  eff}$. The group of 4 cooler stars have a mean [O/H]$_{\rm NLTE}$ that is
  $0.20\pm0.05$\,dex {\it lower} than the 4 stars with $T_{\rm
  eff}>6000$\,K and this is responsible for the high reduced
  chi-squared value in Table~\ref{table-clustermean}.  Blanco 1 is a
  young cluster and its stars are magnetically active as a consequence
  of rapid rotation and dynamo action.  It has been noted by previous
  workers that it can be difficult to obtain oxygen abundances in
  chromospherically active stars using the O\,{\sc i} triplet
  lines. Spuriously high oxygen abundances might be obtained, 
  which are not adequately
  dealt with by NLTE corrections similar to those we adopt here.
  Such abundance over-estimates seem to increase with
  chromospheric activity (Morel \& Micela 2004; Schuler et al. 2004).
  However, both Morel \& Micela and Schuler et al. suggest that this
  effect, (which is possibly attributable to overionisation/excitation
  in the upper atmosphere), {\it does not} seriously compromise oxygen
  abundances determined from the triplet lines for stars with $T_{\rm
  eff}>5500$\,K.  This suggestion needs further testing with high
  signal-to-noise observations of the weak [O{\sc i}]~6300\AA\ feature in
  chromospherically active F and G dwarfs. We note here that the sample
  stars presented in this work are mostly slow rotators (although a low
  inclination angle could mask pole-on rapid rotators), and that the
  derived NLTE oxygen abundances for the cooler stars are {\em lower}
  than for the hotter stars, suggesting chromospheric activity is not
  to blame for the trend we see. In any case it seems that the NLTE
  corrections are not entirely satisfactory and we judge it prudent to
  add a further systematic uncertainty of $\pm 0.1$ dex to the mean
  cluster O abundances.

The Si abundances appear to follow a similar $T_{\rm eff}$ trend to the
O abundances albeit not so significant. Adding a slope to the abundance
versus $T_{\rm eff}$ relationship yields a gradient only 2 sigma above
zero (compared with 4 sigma for O). We note that the strengths and
excitation potentials of the Si lines we have used are similar to,
although higher on average, than those of Mg where a small chi-squared
and no trend with $T_{\rm eff}$ is observed. 
Instead it could be that the Si abundance measurements,
which are the most precise of all the elements considered here,
highlight additional sources of error not yet included (for instance
atmospheric inhomogeneities or a $T_{\rm eff}$-dependent error in the
$T_{\rm eff}$ scale).  If so, it would only have to amount to an
additional 0.03\,dex uncertainty on each stellar Si abundance to reduce the
chi-squared to acceptable ($>10$ per cent probability) levels.  This
level of extra uncertainty would have a negligible impact on the mean
abundances for the other elements.

It is interesting to note that none of the [X/Fe] abundance ratios show any
significant evidence for either trends with $T_{\rm eff}$ or that the
abundance errors are underestimated. This perhaps indicates that any
additional sources of error in [X/H] are mirrored by similar errors in
[Fe/H].

\subsection{External Errors}
\label{exterrors}

Before comparing our Blanco 1 abundances with other published results
for cluster and field stars or using these results as inputs to stellar
evolution calculations, external errors must be considered.  Systematic
shifts in the abundances of our target stars could be caused by
uncertainties in the way we calibrated the $T_{\rm eff}$ scale,
dependence on the atmospheric models, the uncertain age of the cluster
etc.

The sensitivity to
differing atmospheric models of differential abundance analyses on F/G
stars has been investigated by Allende-Prieto et
al. (2004). They found that changing between Kurucz {\sc atlas9} and
{\sc marcs} (Gustafsson et al. 1975) models can alter differential
[X/H] abundances by up to 0.05\,dex. The largest changes were found
amongst abundances derived from lines with high excitation potential,
such as C, O and S.  Likewise, R03 showed that differential [X/H]
abundances altered by only a few hundredths of a dex when changing
between {\sc atlas9} models with and without convective overshoot, but
that differences in [X/Fe] were negligible. R03 also showed that their
[X/Fe] values agreed well (to within a few hundredths of a dex) with
those of Chen et al. (2002) for stars in common, where the latter
authors had used a similar $T_{\rm eff}$ calibration but {\sc marcs}
models. We do not consider these model-dependent uncertainties any
further but the reader should be aware of them.

The effects of a shift in the $T_{\rm eff}$ scale can be estimated by
arbitrarily increasing $T_{\rm eff}$ for all the stars by 100\,K --
which represents a likely level of uncertainty in the ionization
balance calculations and the empirical $T_{\rm eff}$ scales of Alonso
et al. (1996)\footnote{A revision of the Alonso et al. (1996)
  calibrations has been published by Ram\'irez \& Mel\'endez
  (2005). However, their $V-K$--$T_{\rm eff}$ relation yields temperatures only
  20-30\,K hotter for our stars.}. 
The results are listed in Table~\ref{table-clustermean}
for both [X/H] and [X/Fe].  
Such a change results in a mean [Fe/H] change of $+0.04$\,dex and between
$+0.02$ and $+0.09$\,dex for Si, Mg, Ca, Ti, Ni -- hence [Si/Fe] etc.
are changed by less than $\pm$0.05\,dex and [Mg/Fe], [Ni/Fe] by
negligible amounts.  [C/H], [O/H] and [S/H] behave
in the opposite sense, with mean abundances {\em decreasing} by 
0.03 to 0.05\,dex and hence [C/Fe], [O/Fe] and [S/Fe] decrease by 0.08
to 0.09\,dex.

Systematic shifts in the $\log g$ calibration are also possible. If the
cluster was a little older or younger or we were to adopt other
evolutionary models then $\log g$ might change, at most, 
by $\pm 0.2$\,dex. Potentially this could change C abundances by $\pm
0.06$\,dex, O abundances by $\pm 0.04$\,dex and Mg and Ca abundances by
$\mp 0.03$\,dex. Other elements are negligibly affected.

\subsection{Comparison with Edvardsson et al. (1995)}

E95 also calculated mean abundances
(differentially with respect to the Sun) for Blanco~1. Their methods
differed from ours in that they generally used fewer lines and adopted
$T_{\rm eff}$ and $\log g$ values derived from Stromgren photometry
using the calibrations described in Edvardsson et al. (1993).

E95 obtain a mean cluster [Fe/H] (on the basis of four F stars) of 
$+0.23\pm0.01$, [Ni/H] of
$+0.10\pm0.02$, [Si/H] of $+0.13\pm0.05$ and [Ca/H] of $+0.19\pm0.03$.
Their estimated external errors were about $\pm0.1$\,dex.

Our results for [Ni/Fe], [Si/Fe] and [Ca/Fe] match those of E95 reasonably
well. This is reassuring as we claimed in the last section 
that these ratios are quite robust to
changes in models and $T_{\rm eff}$ scales. However, our overall
abundances are lower by 0.18-0.29\,dex, which is several times larger
than our estimates of the internal errors and the external errors
quoted by us or E95.

A comparison of the $\log gf$ values and calibrating solar EWs for
lines in common between the two studies reveal no large or systematic
discrepancies. The root cause of the abundance differences appears to
be in the $T_{\rm eff}$ scale. We have three stars in common with the
E95 study -- W8 ([Fe/H]$_{\rm E95}=+0.23\pm0.02$), W60
([Fe/H]$_{E95}=+0.02\pm0.03$ -- which E95 discount as a cluster member
on the basis of its abundance!) and W63 ([Fe/H]$_{\rm E95}=
+0.21\pm0.01$) -- but in our analysis these stars are cooler by 225\,K,
150\,K and 420\,K respectively. If E95 had adopted the atmospheric
parameters we have used, then their [Fe/H] for W8 and W63 would have been
in excellent agreement with our results. The [Fe/H] of W60 would have
been about 0.15\,dex lower, but E95 explain that the
rotational broadening of this star may have caused them to
underestimate the EWs (and hence abundance), a problem which our
spectral synthesis technique avoids.
 
In deciding what the overall abundances are in Blanco~1 the question of
the adopted $T_{\rm eff}$ scale is crucial. E95 chose to use
temperatures indicated by the Stromgren photometry and the Edvardsson
et al. (1993)
calibration despite the fact that their own spectroscopy -- in the form
of trends of
abundance versus excitation potential and the ionization balance --
indicated that significantly lower temperatures were warranted.  Other
authors (R03) have also noted that the Edvardsson et al. (1993) $T_{\rm
eff}$ scale is 100--150\,K hotter than that deduced from Stromgren
photometry and the Alonso et al. (1996) or 
Saxner \& Hammarback (1985) calibrations. We further note
that for the Blanco 1 stars considered here, and by E95, that
temperatures found from Stromgren photometry and the Alonso et
al. (1996) calibrations are systematically hotter than those based on
$V-K$ by a further $\simeq 150$\,K, 
but are less precise (about $\pm 120$\,K) and are
dependent on metallicity and gravity. For these reasons and
because we are able to achieve ionisation balance at a similar
temperature to that indicated by the $V-K$ photometry, we believe our
abundances are more robust. We do concede that an upward correction to
our $T_{\rm eff}$ scale of $\simeq 100$\,K is still 
possible and the consequences of such a shift were examined
in section~\ref{exterrors}.

\subsection{Comparison with Jeffries \& James 1999}

Four of our cooler targets (W64, W113, ZS58, ZS141) were observed by
JJ99. The temperatures derived there were based on the $B-V$/$T_{\rm
eff}$ calibration of Bohm-Vitense (1981) with an extra
metallicity-dependent term. These temperatures are only 16--116\,K hotter
than the spectroscopically derived temperatures in this paper, but had JJ99
used the metallicities derived here rather than {\em assuming} a
mean cluster metallicity of $+0.14$ then even better agreement would be
obtained.  Only Li abundances were calculated by JJ99, based on curves
of growth presented by Soderblom et al. (1993) and also corrected for
NLTE effects using the code of Carlsson et al. (1994). The difference
between the J99 NLTE Li abundances and those in
Table~\ref{table-results} is $(-0.01\pm0.05)$\,dex.

\section{Discussion}
\subsection{The abundance mix in Blanco 1}

The main results of our analysis are:
\begin{enumerate}
\item The overall metallicity is much lower than found in the previous
  study by E95. This is mainly due to our adoption of a lower $T_{\rm
  eff}$ scale.

\item We confirm the tentative findings of E95, that [Ni/Fe] and
  [Si/Fe] are significantly sub-solar. Furthermore, after considering the
  possible sources of internal and external error, we also find that
  [Mg/Fe] and [Ca/Fe] are sub-solar. The same sub-solar trend is
  indicated for [C/Fe], [S/Fe] and [Ti/Fe] but at a lower level of
  significance. In fact the only solar abundance ratio is that for
  [O/Fe], but the error bar is large enough that it may also be consistent
  with the ratios for the other alpha elements.
\end{enumerate}

The abundance pattern in Blanco~1 is very unusual.  [Ni/Fe], [Mg/Fe],
[Si/Fe] and [Ca/Fe] are all derived from multiple lines with moderate
excitation potentials and line strengths. These ratios are quite robust
to systematic errors and have been measured in a similar way in several
large studies of field dwarfs (Edvardsson et al. 1993; Chen et
al. 2002; R03; Allende-Prieto 2004).  Yet none of these studies contain
{\em any} stars with [Ni/Fe] or [Mg/Fe] as low as we find for Blanco~1,
and stars that are underabundant in the other alpha elements are rare.
On the other hand there is at least one other open cluster, M34, which has
[Ni/Fe]\,$=-0.12\pm0.02$ and [Mg/Fe]\,=$-0.10\pm 0.02$ derived using
spectroscopic methods similar to those used here (Schuler et al. 2003).

The underabundance of both Ni and the alpha elements in Blanco~1 may
perhaps be better considered as an excess of Fe and leaves us with two
puzzles.  The first is that R03 claim that at a
given [Fe/H] there is only a very narrow ($\leq 0.05$\,dex) spread in
[Ni/Fe] and the other abundance ratios discussed here. This is evidence
that at any given time, the ISM in star forming regions of the galactic disc
has been thoroughly mixed and is locally homogeneous.  Blanco~1 is
quite exceptional from this point of view; the gas from which it formed
may not have been well mixed with the bulk of the galactic disc ISM. It
should be noted however that the spectroscopic surveys of field F and G
stars are quite limited in the distances that they probe, although they
span a range of galactic birth site radii of a few kpc.

The second puzzle is the nature of the abundance anomalies in Blanco~1.
The main sources of Fe in the ISM are supernovae of types I and II or
possibly hypernovae.  SN~II and hypernovae invariably produce
super-solar yields of [Si/Fe] (and the other alpha elements). The
relative yields of iron-peak elements are somewhat dependent on the
detailed supernova physics but recent models suggest that the Ni/Fe
ratio is somewhere between 0.75 and 2.0 times the solar value (Nakamura
et al. 2001; Hoffmann, Woosley \& Weaver 2001).  On the other hand
SN~Ia explosions produced by accretion onto a massive white dwarf
produce little Mg but have ejecta with Ni/Fe ratios that are greater than
1.5 times the solar value (e.g. Iwamoto et al. 1999; Travaglio et
al. 2004). It is therefore difficult to understand how the material
from which Blanco~1 formed could appear depleted 
of alpha elements {\it and} Ni with respect to Fe.

The solution to these puzzles (as first noted by E95) could be
connected with a very unusual formation history for Blanco~1. The
cluster lies at high galactic latitude ($b=-79^{\circ}$) and is some
240\,pc below the galactic plane. This is far in excess of the maximum
scale height achieved by similarly young field stellar populations or
open clusters. E95 speculated that Blanco~1 may have formed in the
shocked gas of a high velocity cloud during a collision with the
galactic plane ISM (Comer\'on \& Torra 1994).  In any case, its
undoubtedly peculiar trajectory may mean that the material from which
Blanco~1 originated had travelled some distance and did not have the
chance to homogenise with the local galactic disc ISM.

Perhaps the ISM from which Blanco~1 formed was enriched by one or two
peculiar supernovae? In the literature there are some models of
``delayed detonation'' SN~Ia, produced by the merger of two white
dwarfs, which do produce more Fe than Ni (Khokhlov, M\"uller \&
H\"offlich 1993). If, as suggested by E95, Blanco~1 formed from a
3000\,M$_{\odot}$ cloud, then a single such event could pollute the
cloud sufficiently to raise [Fe/H] by 0.1\,dex, but would only decrease
[Ni/Fe] by 0.03\,dex, which does not seem sufficient.

\subsection{Implications for Li depletion}

Li in Blanco~1 was investigated by JJ99 who wished to determine whether
its higher metallicity with respect to (for instance) the Pleiades
would result in increased levels of PMS Li burning and
depletion, as predicted by standard evolutionary models. That no
significant differences were seen in the Li-depletion patterns of
Blanco~1 and the Pleiades indicated problems for the standard models
and that a mechanism which inhibited Li-depletion in Blanco~1 was
required.

JJ99 assumed a mean metallicity [M/H] of $+0.14$ for Blanco~1.  In this
paper we have determined a lower value; furthermore we have
found that if anything, the abundances of other alpha elements which
are an important source of radiative opacity in the envelopes of PMS
stars, primarily O, Mg and Si (Piau \& Turck-Chi\'eze 2002) are even
lower. The Pleiades has [Fe/H]\,$=-0.034\pm0.024$ (Boesgaard \& Friel
1990), [O/H]\,$=+0.02\pm0.10$ (based on the O\,{\sc i} triplet
measurements of stars with $T_{\rm eff}>5500$\,K in Schuler et al. 2004
and applying the same NLTE corrections as in this work),
[Si/Fe]\,$=+0.09$, [Ca/Fe]\,$=+0.07$ and [Ni/Fe]\,$=+0.02$ (Wilden et al. 2002). Thus the
mean metallicity and hence radiative opacities in Blanco~1 are probably
quite similar to those in the Pleiades.

The predictions of standard models are therefore adequate
to explain why the Li depletion patterns of cool stars in the Pleiades and
Blanco~1 are indistinguishable. However the same models do have other
problems (they deplete too much Li overall and cannot explain the
scatter in Li abundances for young cluster K-stars -- e.g. Jeffries
2005) and there are other sets of clusters with similar ages that show
almost identical levels of Li-depletion despite their differing
(spectroscopically determined) [Fe/H] values (e.g. Jeffries et al. 2002;
Sestito, Randich \& Pallavicini 2004).
A thorough spectroscopic investigation of the individual
elemental abundances of these clusters is certainly warranted.

\section{Summary}

We have determined elemental abundances for a group of eight F- and
G-type stars that are good candidate members of the young open cluster
Blanco~1. The mean cluster [Fe/H] is $0.04\pm 0.02 \pm
0.04$\footnote{Here and below, 
the first uncertainty is an internal error
  incorporating statistical errors and uncertainties in atmospheric
  parameters for each star. The second uncertainty is a systematic
  external error and incorporates
  possible errors of 100\,K in the adopted temperature scale and
  0.2 in the adopted $\log g$ values 
(see section~\ref{exterrors}).}, considerably lower
than a previous spectroscopic determination by Edvardsson et
al. (1995). The difference is mainly due to the adoption of a
significantly cooler $T_{\rm eff}$ scale in this paper, which we argue
is more consistent with both photometric and spectroscopic indicators.

We find that Blanco~1 has significantly subsolar values of [Si/Fe]
($=-0.09\pm0.02 \pm 0.03$) and [Ni/Fe] ($=-0.18\pm0.01 \pm 0.01$) and
extend this trend to include [Mg/Fe] ($=-0.14\pm0.02 \mp 0.03$),
[Ca/Fe] ($=-0.09\pm0.02 \pm 0.03$) and [Ti/Fe] ($=-0.10\pm0.03
\pm0.05$).  The abundance ratios of [O/Fe], [C/Fe] and [S/Fe] are
derived from high excitation potential lines and subject to larger
uncertainties. They may be consistent with the underabundances seen for
the other elements but could also assume solar values.

The combination of a deficit of both Ni and the alpha elements Mg and
Si with respect to Fe is difficult to explain with published
models of SN~Ia and SN~II. These subsolar abundance ratios are also
very unusual when compared with samples of nearby stars that have had
their abundances measured using similar techniques. This may indicate
that the material from which Blanco~1 formed was not well mixed with the
ISM of the galactic disc and may have been polluted by one or two
unusual supernova events. The large distance of Blanco~1 from the
galactic disc combined with its youth also point to an unusual 
formation history.

Our revised determination of the Blanco~1 abundances could explain why
the Li depletion pattern observed among its G- and K-type members is
indistinguishable from those in the Pleiades. The overall radiative
opacities and hence convection zone properties at a given mass will be
very similar in the two clusters.  Unfortunately, our lower revised
abundances for Blanco~1 also mean it may not be such a fruitful place
to look for short-period exoplanets.

\section*{Acknowledgments}

This research has made use of NASA's Astrophysics Data System, and the
SIMBAD Astronomical database and VizieR service hosted by CDS. We thank
the director and staff of the Anglo Australian Observatory for
assisting with the success of our observations.

\bibliographystyle{mn2e}  
\bibliography{iau_journals,master}

%%%%%%%%%%%%%%%%%%%%%%%%%%%%%%%%%%%%

\bsp % ``This paper has been produced using the ...''

\label{lastpage}

\end{document}